\documentclass[12pt,english,a4paper]{article}
\usepackage{natbib}
\usepackage{rotating,placeins}
\usepackage{babel}
\usepackage{geometry}
\usepackage[latin1]{inputenc}
\usepackage[T1]{fontenc}
\usepackage{graphicx,graphics,mathpple,textcomp}
\usepackage{fancyhdr}
\usepackage{lscape,epsfig,amssymb,amsmath}
\usepackage[]{subfigure}
\usepackage{rotating,placeins}
\usepackage{lscape}
\usepackage{setspace}
\newcommand{\Sec}[1]{Sec~\ref{sec:#1}}
\newcommand{\Fig}[1]{Fig~\ref{fig:#1}}
\newcommand{\Eq}[1]{Eq~(\ref{eq:#1})}

\title{Wave Extremes in the North East Atlantic from Ensemble Forecasts}
\author{{\O}yvind Breivik\footnote{Final version published as Breivik, {\O}, O~J Aarnes, J-R Bidlot, 
 A Carrasco and {\O} Saetra (2013). Wave Extremes in the North East Atlantic from Ensemble Forecasts, 
 \emph{J Climate}, doi:10.1175/JCLI-D-12-00738.1 (in press)}
\thanks{Corresponding author. E-mail: \texttt{oyvind.breivik@ecmwf.int}}
\thanks{European Centre for Medium-Range Weather Forecasts (ECMWF), Shinfield Park, Reading, RG2 9AX, 
 United Kingdom}, 
\and Ole Johan Aarnes\thanks{Norwegian Meteorological Institute}
\and Jean-Raymond Bidlot\footnotemark[3]
\and Ana Carrasco\footnotemark[4]
\and {\O}yvind Saetra\footnotemark[4]}

\begin{document}
\maketitle                                                                                                          

\abstract{
A method for estimating return values from ensembles of forecasts at advanced
lead times is presented.  Return values of significant wave height in the
North-East Atlantic, the Norwegian Sea and the North Sea are computed from
archived +240-h forecasts of the ECMWF ensemble prediction system (EPS)
from 1999 to 2009.

We make three assumptions: First, each forecast is representative of a six-hour
interval and collectively the data set is then comparable to a time period
of 226 years. Second, the model climate matches the observed distribution,
which we confirm by comparing with buoy data. Third, the ensemble members
are sufficiently uncorrelated to be considered independent realizations of
the model climate.  We find anomaly correlations of 0.20, but peak events
($>P_{97}$) are entirely uncorrelated. By comparing return values from
individual members with return values of subsamples of the data
set we also find that the estimates follow the same distribution and appear
unaffected by correlations in the ensemble.  The annual mean and variance
over the 11-year archived period exhibit no significant departures from
stationarity compared with a recent reforecast, i.e., there is no spurious
trend due to model upgrades.

EPS yields significantly higher return values than ERA-40 and ERA-Interim
and is in good agreement with the high-resolution hindcast NORA10, except in
the lee of unresolved islands where EPS overestimates and in enclosed seas
where it is biased low.  Confidence intervals are half the width of those
found for ERA-Interim due to the magnitude of the data set.  }

\newpage
\section{Introduction}
\label{sec:intro}
Extreme value estimates of atmospheric and oceanographic variables are usually
derived from observational records or from model reconstructions of the past
(reanalyses and hindcasts).

A given probability of exceedance or equivalently return period corresponds
to a return value of the geophysical variable in question. This return
value is normally approximated by fitting the Generalized Extreme Value
(GEV) distribution to blocked maxima (such as annual maxima) or by fitting
the Generalized Pareto (GP) distribution to exceedances above a threshold,
see \cite{smi90b} and \cite{col01}.  In the atmospheric and oceanographic
sciences 100-year return values are usually sought \citep{lop00}, which
means that observational or modeled time series are rarely long enough to
cover the return period, even for the longest reanalyses and hindcasts (see
\citealt{upp05,kalnay96,wang12} for descriptions of some recent reanalyses).
Extrapolation of the parametric fit to lower probabilities of exceedance
(return periods longer than the observational record or modeled time series)
is then required. This affects the confidence intervals of the return
value estimates and is a concern when using shorter records like altimeter
measurements \citep{alves03,you11}.  Model or observational bias will further
increase the confidence intervals, but is much harder to identify than the
unsystematic error stemming from insufficient length of the time series.

Trends and low-frequency oscillations can seriously influence return
value estimates from time series. This can be handled using non-stationary
techniques (see \citealt{col01}, Chapter 6 for an introduction). Due to
imminent climate change \citep{ipcc07}, estimating return values from
time series with trends has recently received some attention in the earth
sciences. \cite{kharin00} and \cite{kharin05} investigated the impact
of a linear trend on the GEV distribution of the annual extremes while
\cite{parey07} looked at extending the extreme value theory to assess the
return values of temperature extremes in the presence of a linear trend over
a 54-year period for French observing stations.  \cite{winter12} investigated
the changing wave extremes in a regional climate projection of the North
Sea for the time-slice 2071-2100. Similarly, \cite{wang04} and \cite{wang06}
investigated the impact of changing wave climate on wave extremes in the span
of the 21st century using statistical projections and coupled climate models.

Even if non-stationarity can be handled it raises the question of what
exactly the return value estimates are to be used for. If a probability
of exceedance valid for a certain time period is required, similar to what
\citealt{kharin00} did for 21-year time slices from climate projections
considered sufficiently stationary, then a long time series is not necessarily
of much interest. What is then needed is an estimate of exceedance levels for
that given time slice. Such a repository of possible weather realizations
does in fact exist. The ensemble prediction system (EPS) operated by the
European Centre for Medium-Range Weather Forecasts (ECMWF) has now been in
operation for 20 years \citep{mol96,buizza07,hagedorn08}.  The individual
ensemble members start from almost identical initial conditions with only small
perturbations added to the best guess analysis \citep{buizza99,leutbecher08}
to spread the ensemble in a way representative of the uncertainty of
the forecast system. Although there is considerable forecast skill after a lead
time of five days \citep{richardson10}, the skill drops rapidly after day
six, and on day 10 the individual members are only weakly correlated with
each other and with observations, as we will show in \Sec{data}. If the
quantiles of the entire cumulative distribution of the ensemble compare well
with observations then the forecasts can be considered random realizations
of a realistic model climate.

Return values for significant wave height have been estimated from
a wide variety of data sources in the past, ranging from relatively
short observational records (\citealt{battjes72} and \citealt{muir86}),
satellite altimeters \citep{cooper97,alves03,vinoth11}, long-term
global reanalyses \citep{caires05b,sterl05}, regional model hindcasts
\citep{wang01,wang02,weisse07,aar12}, and statistical downscaling
\citep{bre09}. Here we explore a new approach to estimating return values of
significant wave height using ensemble forecasts at advanced lead times instead
of a time series.  A similar approach has been explored by \cite{brink05} for
the special case of river flooding protection using seasonal forecast ensembles
from ECMWF's earlier System 2 Seasonal Forecast System \citep{anderson03}.
\cite{brink05} employed the entire seasonal forecasts from a lead time of one
month up to six months, arguing that the modelled North Atlantic Oscillation
(NAO) was only weakly correlated with observed NAO after one month, dropping
further to essentially zero for the subsequent months. We employ a different
approach where we instead extract the significant wave height for a fixed
forecast time (+240 h) from the EPS version of the Integrated Forecast
System (IFS) of ECMWF. We have gathered all forecasts at +240~h generated
during the period 1999-2009, equivalent (as will be explained in \Sec{data})
to ${\sim}226$ years if the data had formed a continuous time series. As
will be explained in \Sec{method} we assume that each forecast represents a
six-hour interval, which is a reasonable assumption for a coarse model and
analogous to the temporal resolution of traditional reanalyses. However,
this also means that we estimate return values of the six-hourly average
sea state.  We address this in \Sec{method} and discuss the implications
further in \Sec{discussion}.

The method to be explored allows us to utilize a vast unused resource of
climate realizations and their lack of skill is actually a prerequisite since
extreme value theory demands that events be uncorrelated. However, there
are important caveats to the interpretation and use of the method. First,
climate trends are by construct not captured by the method since we base our
estimates on a time-slice of ${\sim}10$ years. Likewise, quasi-cyclical
phenomena like El Ni\~{n}o with a period longer than what is covered by
the archive may influence the results.  This suggests the following use
and interpretation of EPS return values: If probabilities of exceedance for
the present time period are sought, then the ensemble data set is superior
since it is not affected by long-term trends and low-frequency cycles.
If on the other hand long-term return values are required then techniques
for estimating extremes from time series with trends must be considered
(see \citealt{kharin00,kharin05,parey07}), or at least comparison with
traditional time series covering a sufficiently long period.

The paper is organized as follows. \Sec{data} presents the observational
records and the reanalysis and hindcast data sets used to test the method.
\Sec{method} presents the method used to compute return values from forecast
ensembles at long lead times and how it differs from traditional return
value estimates from observational records and modeled time series.  We then
investigate the independence of ensemble members and the climatology of the
archived forecasts by comparing against a model climatology from a recent
reforecast (see \citealt{hagedorn12}) and observations.  \Sec{results}
compares the return values found from the EPS with three reference model
data sets, namely the reanalyses ERA-40 and ERA-Interim (ERA-I hereafter)
and a high-resolution regional hindcast for the Norwegian Sea and adjacent
seas, NORA10 (see \citealt{rei11,aar12}).  \Sec{discussion} discusses the
differences in method and results, and points at possible weaknesses of the
method. Finally, \Sec{conclusion} presents our conclusions on the general
usefulness of the method and its application to significant wave height and
the ECMWF EPS system.

\section{Modelled and Observed Wave Climate}
\label{sec:data}
To assess the validity of our return value estimates from EPS forecasts, we
will make a number of comparisons with observational records, reanalyses
and hindcasts of significant wave height. This section presents the
observations used and the five model data sets (ERA-40, ERA-I, NORA10,
EPS and EPS reforecasts). We investigate the EPS climatology at analysis
time (labeled EPS0) and at +240-h lead time (labeled EPS240) and assess the
stationarity and independence of EPS240. Time series of all model data have
been interpolated to buoy locations.

Time series have been extracted from ERA-40, ERA-I, NORA10 and EPS and
interpolated to the same $1^\circ \times 1^\circ$ grid of the northeastern
part of the Atlantic Ocean, the North Sea and the Norwegian Sea in order to
make geographical comparisons of extreme value estimates. The regridding
and interpolation will inevitably smooth the field slightly. This will
influence the return values somewhat.  It is of interest to compare our EPS
return estimates with these reference data sets because all three archives
(ERA-40, ERA-I and NORA10) are frequently used for return value estimation
(see e.g. \citealt{caires05b,aar12}).

\subsection{ERA-40}
Significant wave height from the ERA-40 reanalysis \citep{upp05}
is available for the period September 1957 to August 2002 on six-hourly
temporal resolution. The atmospheric model was coupled to a deep-water
version of the wave model (WAM) through exchange of a wave-modified Charnock
parameter (\citealt{janssen02} and \citealt{jan04} pp 232--234). WAM was run
on a regular $1.5^\circ$-grid.  At this resolution the Shetland and Faroe
archipelagoes are not resolved and the modeled wave field on the lee side
of these islands is consequently biased high. It is also well known that
ERA-40 is biased low in general \citep{cai05,rei11}.  For this study we have
not attempted any correction either to the time series themselves, which
is how \cite{cai05} came up with the corrected semi-global fields referred
to as the Corrected ERA-40, or by correction of the 100-year return values,
which is how \cite{caires05b} and \cite{sterl05} constructed the global maps
of return values from ERA-40. We argue that for this study it is better to
compare the original data sets to avoid confounding artefacts of the new
approach with artefacts of the statistical correction algorithms employed
by \cite{caires05b} and \cite{cai05}. However, we do discuss how our results
qualitatively correspond to the results of \cite{caires05b} in \Sec{results}.

\subsection{ERA-Interim}
ERA-I is a continually updated coupled atmosphere-wave reanalysis
which originally covered the period from 1989 (roughly co-incident with
the satellite era), but which has recently been extended back to 1979
\citep{simmons07,uppala08,dee11}. The resolution is $1.0^\circ$ for the wave
model at the equator, but the resolution is kept nearly constant towards
the poles by the use of an irregular latitude-longitude grid.  The wave
model is coupled to the atmospheric model in the same fashion as outlined
above for ERA-40, but the ERA-I wave model physics include shallow-water
effects important in areas like the southern North Sea. ERA-I also differs
from ERA-40 in its use of a four-dimensional variational assimilation scheme
and a substantially larger amount of observations, especially after 1991.
ERA-I uses a subgrid scheme to represent the downstream impact of unresolved
islands \citep{bidlot12}. Though a clear improvement over ERA-40, the wave
field in the lee of the Faroes and the Shetland Isles is still biased a
little high.

\subsection{The NORA10 regional hindcast} 
NORA10, a recently completed atmospheric downscaling of ERA-40 and wave model
hindcast on 10-11 km resolution, is described by \cite{rei11}. The model domain
covers the North Sea, the Norwegian Sea and the Barent Sea. The temporal
resolution of the archived fields is three hours. The hindcast initially
covered the ERA-40 period (September 1957 to August 2002), but an extension
with boundary and initial fields from the ECMWF IFS has since been added.
The hindcast archive is continually updated. The breach of stationarity due to
the change in boundary and initial values after August 2002 was investigated by
\cite{aar12} and no statistically significant changes were found.  The median
and upper percentiles of NORA10 $H_\mathrm{s}$ show little bias and generally
close correspondence with the wave observations used in this study. The model
resolves the main coastal features and the archipelagoes in the Norwegian
Sea. Like ERA-I the wave model is run in shallow water mode.

\subsection{ECMWF EPS archive}
We have extracted the significant wave height from archived operational
ECMWF EPS wave forecasts for the period 1999-2009, a total of 11 years. The
data set is not homogeneous, i.e., the resolution and the model physics
of the operational EPS forecast system have been continually upgraded
(see \Fig{timeline} for the most important changes affecting the wave
field). The wave model has been coupled to the atmospheric model in the
same fashion as for ERA-I.  The data assimilation scheme has been upgraded
several times during the archived period, and the amount of data entering the
assimilation cycle has steadily increased. It is also important to note that
the forecast systems started issuing two forecasts per day on 2003-03-25 (00
and 12 UTC analysis time). This means that the amount of data is not uniform
over the period. We have extracted the analysis and the +240-h forecasts
from the 50 perturbed ensemble members plus the control member (forced by
unperturbed wind fields).  We have also extracted the forecasts at +228 h
(EPS228 hereafter). This data set is naturally slightly more correlated than
EPS240 and is used here primarily to assess the validity of the method.

\subsection{EPS reforecasts}
\label{sec:reforecasts}
Since 2008 every new model cycle of the EPS has been accompanied by a model
climatology based on reforecasting a five-member ensemble of the current
model cycle (four perturbed and one unperturbed member) from ERA-I initial
conditions every Thursday from 18 years back and up until the present day
\citep{hagedorn08,hagedorn08b,hagedorn12,prates11}.  These reforecasts are
run to 32 days, similar to the operational forecasts.  We have extracted the
reforecasts valid at +240 h from model cycle Cy36r4, in operation after the
end of the archived period, from 60 locations in the north-east Atlantic,
the Norwegian Sea and the North Sea. The reforecasts are in principle also
useful for extreme value analysis, but unfortunately the data set is too
small by two orders of magnitude to allow the kind of analysis attempted
with the EPS forecasts:
\begin{equation}
   \frac {5\, \mathrm{weekly\,members} \times\,18\, \mathrm{yr}} 
    {51\, \mathrm{members} \times\,2\, \mathrm{daily\,forecasts} \times 7\,
    \mathrm{weekdays} \times\,11\, \mathrm{yr}} \approx 0.011.
\end{equation}

\subsection{Comparing observed and modeled significant wave height}
\label{sec:obs}
Wave observations are routinely archived and quality-controlled by ECMWF as
part of the wave model intercomparison effort \citep{bidlot02}.  To make the
observations comparable with model output \cite{bidlot02} averaged observations
over four hours centered on the synoptic times. The rationale behind this
averaging procedure is as follows.  Typical wind conditions in the open ocean
are on the order of $10\,\mathrm{m}\,\mathrm{s}^{-1}$. For fully developed
wind sea the group speed, which dictates the propagation speed across the
model grid, will be comparable to the wind speed \citep{holthuijsen07,wmo98}.
If the resolution is ${\sim}1.5^\circ$ then the time interval that is
represented by the model output is 4-6~h. Archived model values, although
``instantaneous'' in the sense that they are model output, are thus slowly
changing and should be considered averages representative of intervals of
4-6~h in the case of the coarser archives discussed below (ERA-40, ERA-I
and EPS). The NORA10 archive has much higher resolution (10-11 km) and is
consequently also archived at three-hourly resolution.  Both ERA-40 and ERA-I
are archived on six-hourly resolution and the return values derived from these
reanalyses should be interpreted as six-hourly averages of the significant
wave height. EPS is of comparable resolution and we assign the same interval
to the EPS240 forecasts.  We discuss this further in \Sec{discussion}.

Three locations with quite different wave climate were selected from a total
of 60 observation stations in the North East Atlantic, The Norwegian Sea and
the North Sea for inspection of the relative performance of the reanalyses
and the EPS:
\begin{itemize}
 \item P40 (Ekofisk oil field, WMO code LF5U, 56.50$^\circ$ N, 003.20$^\circ$ E) in the central 
  North Sea
 \item P35 (Heidrun oil field, WMO LF3N, 65.30$^\circ$ N, 007.30$^\circ$ E) in the eastern
 Norwegian Sea
 \item B16 (K5 buoy, WMO 64045, 59.10$^\circ$ N, 011.40$^\circ$ W) in the north-east Atlantic
\end{itemize}

\section{Estimating return values from ensemble forecasts}
\label{sec:method}

\subsection{Extreme value distributions applied to ensemble forecasts}
\label{sec:evd}
The two most commonly used statistical methods for estimating return values
are based on the GEV distribution for blocked maxima and the GP distribution
for values exceeding a set threshold.  For completeness we repeat here the
general form of the GEV and the GP distributions but refer to \cite{col01} for
a more in-depth discussion.  The GEV distribution (GEVD) is an asymptotic limit
for a distribution of blocked maxima $M_n = \max\{X_1, X_2, \ldots, X_n\}$.
The method is routinely used to approximate the probability distribution of
blocked maxima such as annual maxima (\citealt{col01}, pp 45--51).  However,
the method can also be used on ensembles of independent and identically
distributed (iid) forecasts since the blocking procedure itself makes no
assumption of the grouping other than to ensure that all blocks have the same
statistical properties. The blocking can thus be performed in many ways,
but it is natural to block by ensemble member or some subset of time and
member which is sufficiently large to ensure that the GEVD is a reasonably
good approximation to the parent distribution. Following \cite{col01} the
cumulative distribution function (CDF) of the block maxima formed from a
random sequence of independent variables can be written
\begin{equation}
   G(z) = \exp \left\{ -\left[1+\xi
   \left(\frac{z-\mu_n}{\sigma_n}\right)\right]^{-1/\xi}\right\},
   \label{eq:gev}
\end{equation} 
where $\sigma_n$ is the scale parameter, $\mu_n$ is known as the location
parameter, and $\xi$ is the so-called shape parameter. The GEV distribution
contains as special cases the Fr\'{e}chet ($\xi > 0$), Gumbel ($\xi = 0$)
and reversed Weibull ($\xi < 0$) distributions.  The width of confidence
intervals will depend strongly on the sign of the shape parameter
\citep{hosking84,col01}.

The GP method retains only values exceeding a threshold $u$.  The transformed
variable is written $y = X_i - u$, $y > 0$. It can be shown (\citealt{col01},
pp 75--77) that the GP distribution (GPD) is applicable if the data $y$ are
independent and the maxima $M_n$ formed from the original variable belong
to the GEV distribution, \Eq{gev}. The GDP is written
\begin{equation}
   H(y) = 1 - \left(1+\frac{\xi y}{\tilde{\sigma}}^{-1/\xi}\right).
   \label{eq:gp}
\end{equation}
Here $\tilde{\sigma} = \sigma + \xi(u-\mu)$ and $\xi$ is the shape parameter
found in \Eq{gev}. For $\xi = 0$ GP becomes an exponential distribution.
In traditional studies of wind and wave extremes it is common to select only
peaks separated by at least 24-48 hours to ensure that the maxima represent
individual storm events \citep{lop00,aar12,nae01}.  This is known as the
``Peaks-over-threshold'' method (POT) and it is the method we apply to
the ERA-40, ERA-I and NORA10 time series.  Since we assume (see below)
that the EPS forecasts are uncorrelated at +240 h the ensemble members
represent independent events and we can retain all values exceeding the
chosen threshold. In this case the return values are more properly referred
to as GP threshold estimates rather than GP/POT estimates.

\subsection{Criteria for using ensembles for extreme value estimation}
\label{sec:criteria}
The following assumptions must be shown to hold in order to estimate return 
values from ensemble forecasts:
\begin{enumerate}
  \item Each forecast is representative of a time interval (e.g. six hours) 
  \item The model climatology distribution is comparable to the observed climatology
  distribution
  \item No spurious trend in the mean and the variance 
  due to model updates
  \item No significant correlation between ensemble members at advanced lead 
  times 
\end{enumerate}
If these conditions are met the ensemble can be assumed to be independent
and identically distributed and representative of the observed wave climate.
We will address each of these conditions below.

\subsubsection{Estimating return periods from ensembles}
\label{sec:eps2ts}
Turning $M$ ensemble forecasts with $N$ ensemble members each into the
equivalent of a time period is necessary in order to convert from probability
of exceedance to a return period. We assume each forecast to represent
a six-hour interval based on the following reasoning.  First, $\Delta t =
6\,\mathrm{h}$ matches the temporal resolution of the ERA-I and ERA-40
archives, simplifying the comparison. Second, as discussed in
\Sec{data}(\ref{sec:obs}),
model fields are smoothly varying in time, making them representative
of averages over typically 4-6 hours at the resolution of ERA-40, ERA-I
and EPS.  This allows us to treat the collection of ensemble forecasts as
an \emph{equivalent time period} $T_\mathrm{eq} = M N \Delta t$.  We discuss
the validity of this assumption in \Sec{discussion}.

\subsubsection{Climatology of ensemble forecasts at advanced lead times}
To convince ourselves that the EPS240 dataset is identically distributed we
need only look at the quantile-quantile (QQ) distribution of two members.
Panel (c) of \Fig{corr} clearly demonstrates the similarity of the
distributions of two randomly selected ensemble members (the statistics
look essentially the same for all combinations of members, not shown). The
QQ-plot against observations in panel (a) of \Fig{corr} makes it clear that
the ensemble members at +240 h represent the observed climate well. In fact,
the cumulative distribution of the ensemble at +240 h is somewhat better
aligned with observations than at analysis time as can be seen in panel (b)
of \Fig{corr} for the Ekofisk location in the North Sea, but the differences
in distribution are small from analysis time to +240 h for all the observation
locations and we conclude that we can assume that the ensemble members are
identically distributed and that they represent the observed wave climate well.

\subsubsection{Stationarity of model climate}
To address the question of whether there is a spurious trend in the
model climate over the archived period we compare against the reforecasts
for the same period from Cy36r4. We are only interested in the behavior of
the +240 h forecasts as our objective is to investigate whether the changes
in model physics and resolution, especially in the early days of the EPS
forecast system, will have significant impact on our return value estimates.

The EPS forecasts at lead time +240 h are compared with the reforecasts at
the same lead time and at the same analysis dates over the period 1999-2009
in \Fig{reforecast}.  Panel (a) presents the time series of the annual mean
deviations of the significant wave height at 60 locations in the north-east
Atlantic, the North Sea and the Norwegian Sea. The annual
mean difference in their standard deviation is shown in panel (b).  It is
clear from the boxplots that there is considerable deviation in both mean and
standard deviation between the EPS forecasts and the EPS reforecasts throughout
the model domain from one year to another, but there is no consistent drift
in either statistic. However, it is clear that the reforecasts have slightly
higher mean (${\sim}0.05$ m) and standard deviation (${\sim}0.10$~m).

\subsubsection{Independence of ensemble forecasts at advanced lead times}
\label{sec:iid}
The motivation for using ensemble forecasts at long lead times is the
anticipation that the upper percentile of the data will be found to be
independent. In other words, we are looking for forecasts with as little
skill as possible \citep{wil06}. With a set of ensemble forecasts at advanced
lead time (e.g. 240 hours), the question is then whether the correlation
between two arbitrary ensemble members is sufficiently low for the members
to be assumed independent.  The residual correlation (after subtraction
of the seasonal mean) between two fields can be investigated through the
centered anomaly correlation (\citealt{wil06}, pp 311--312) which for two 
ensemble members $i$ and $j$ can be defined as
\begin{equation}
    r_\mathrm{ACC} = \frac{\mathrm{Cov}(x'_i, x'_j)}{s'^2},
\end{equation}
where primes indicate centered anomalies.  The anomaly correlation at all
three locations is approximately $0.20$, while the actual correlation varies
from $0.46$ in the open ocean (B16 and P35) to $0.35$ at Ekofisk (P40) in
the central North Sea. With small variations this is the level of anomaly
correlations found throughout the domain studied here.

Even weak correlations between ensemble members (if significant) may have
a deleterious effect on the equivalent or effective ensemble size. This is
a well known problem often discussed in the context of autocorrelated time
series (\citealt{sto99b}, pp 371--372).  Denote the sample size (in our case
the number of archived forecasts) $M$ and the ensemble size $N$. The entire
ensemble is $X \in \mathbb{R}^{M \times N}$.  The ensemble variance-covariance
matrix is written $\langle e_i e_j \rangle \in \mathbb{R}^{N\times N}$,
where $e_i$ represent departures from the ensemble mean.  If we assume all
members to have equal variance $\langle e_i e_i \rangle = s^2$ and common
correlation $r$ (a reasonable assumption since there is nothing to distinguish
one member from another) such that $\langle e_i e_j \rangle = rs^2$ (where
we note that $r \equiv 1$ when $i = j$) we arrive at the following relation
for the variance of the ensemble mean,
\begin{equation}
   s^2_{\overline{x}} = \frac{s^2}{N}\left(1 + (N-1)r\right).
\end{equation}
The effective sample size is now found from $s^2_{\overline{x}} = s^2/N^*$,
i.e., the variance of the mean of a smaller ensemble of uncorrelated members
should equal that of our correlated ensemble. The effective ensemble size
becomes
\begin{equation}
   N^* = \frac{N}{1 + (N-1)r}
\end{equation}
and it is clear that even quite weak correlations can seriously reduce the
effective ensemble size and have a detrimental impact on the mean properties of
the ensemble.  However, assessing the impact of correlations on the ensemble
mean alone is of limited value since only the upper percentiles of the data
set are actually used for the return value estimates. To investigate the
possible impact of correlations on the tail of the data set we first computed
the correlation and the Spearman rank correlation (see \citealt{pre92}) for
a subset of the forecasts where \emph{at least one} ensemble member exceeded
the 97 percentile ($P_{97}$). Members not exceeding the threshold were set
to zero. The average rank correlation and Pearson's correlation coefficients
were 0.05 for this subset of forecasts. This shows that the higher percentiles
of the ensemble tend to be uncorrelated even if the ensemble itself exhibits
weak correlation. This is not surprising given the nature of our analysis. We
are selecting the upper percentiles from a large data set. This means that
we are only selecting storm events, which are transient and fast-moving. It
is unlikely that storm events exceeding $P_{97}$ will occur simultaneously
in many ensemble members after a 10-day integration.  Average sea
state will on the other hand be more correlated at long lead times since
such weather patterns are less transient (e.g., high pressure situations).

To assess the impact of any residual correlation on the return values we
followed a heuristic approach suggested by M. Leutbecher (pers comm) where
return values from $N$ individual ensemble members are compared with return
values from decimated subsets of similar sample size where all members are
used, but only every $N$-th forecast. We thus arrive at two distributions
of return values drawn from samples of the same size $M$. Splitting the
total data set in $N$ parts obviously increases the uncertainty associated
with the return value estimates, but we are only interested in comparing
the distributions of the two sets of return values.  \Fig{h100} compares
quantiles of the 51 member-wise return values (first axis) with the 51
return values from the decimated samples for location P40 in the central
North Sea.  It is evident that the two distributions are very similar with
only slightly higher standard deviation (1.83~m v 1.67~m) for the member-wise
estimates. The average is practically the same (11.30~m). We conclude that
the weak correlations found between ensemble members in the mean have no
discernible impact on the expected value or the spread of the return estimates.

\section{Comparison of extreme value estimates and their confidence intervals}
\label{sec:results}
\begin{figure}
\begin{center}
\hspace{5mm}
\includegraphics[scale=0.6]{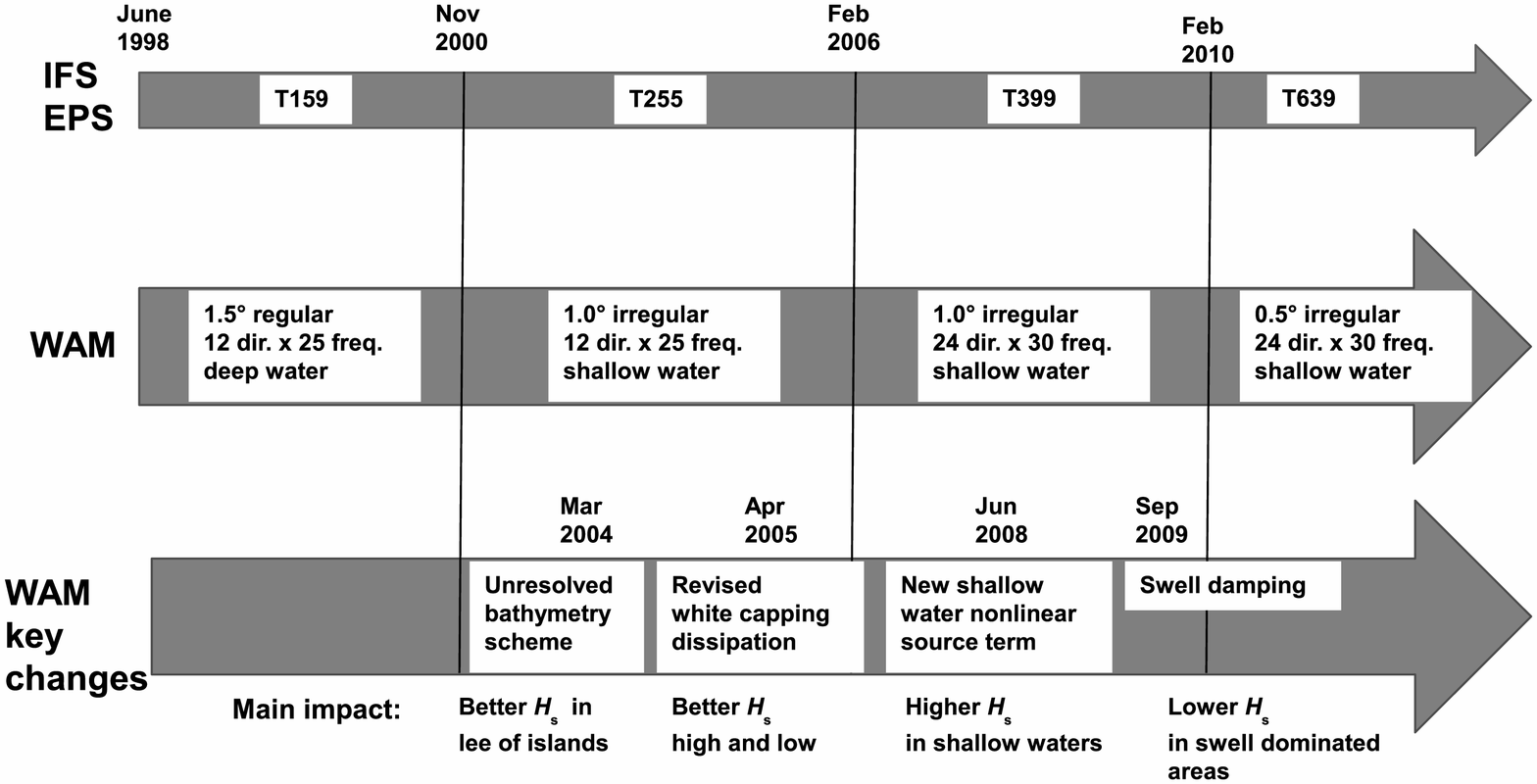}\\
\caption{
A timeline of major updates to the operational EPS. The upper arrow illustrates
changes to the atmospheric component of the integrated forecast system (IFS)
while the middle arrow lists the major changes to the wave model (WAM). Note
that only one forecast per day (00 UTC) was issued before 2003-03-25. The
most important changes related to the wave model are the changes in resolution
which affect the areas in the lee of the Faroes and the introduction of
shallow-water physics which mainly affects the southern North Sea.}
\label{fig:timeline}
\end{center}
\end{figure}

\begin{figure}
\begin{center}
\begin{tabular}{cc}
(a)\includegraphics[scale=0.45]{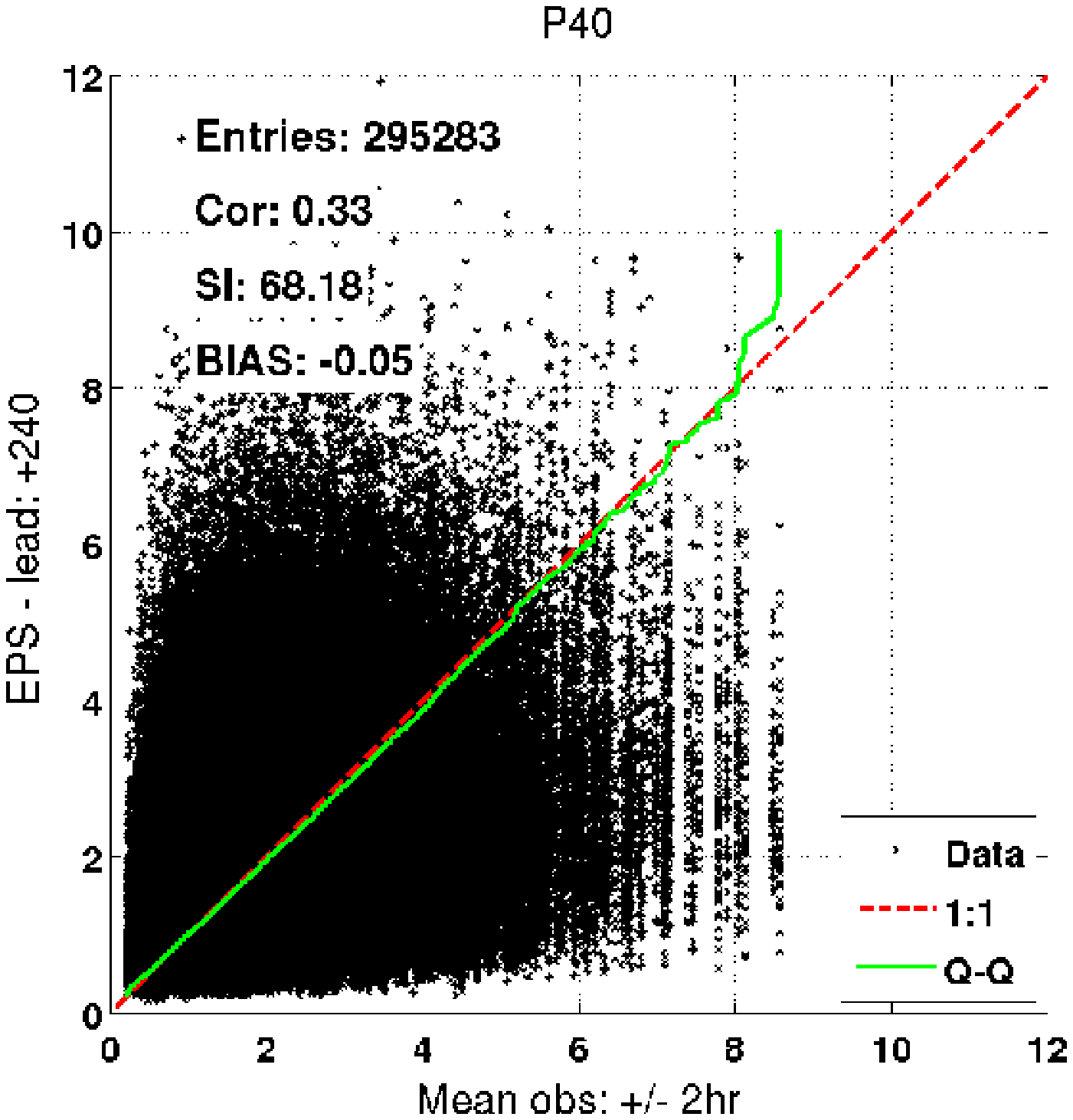}&
\\
(b)\includegraphics[scale=0.45]{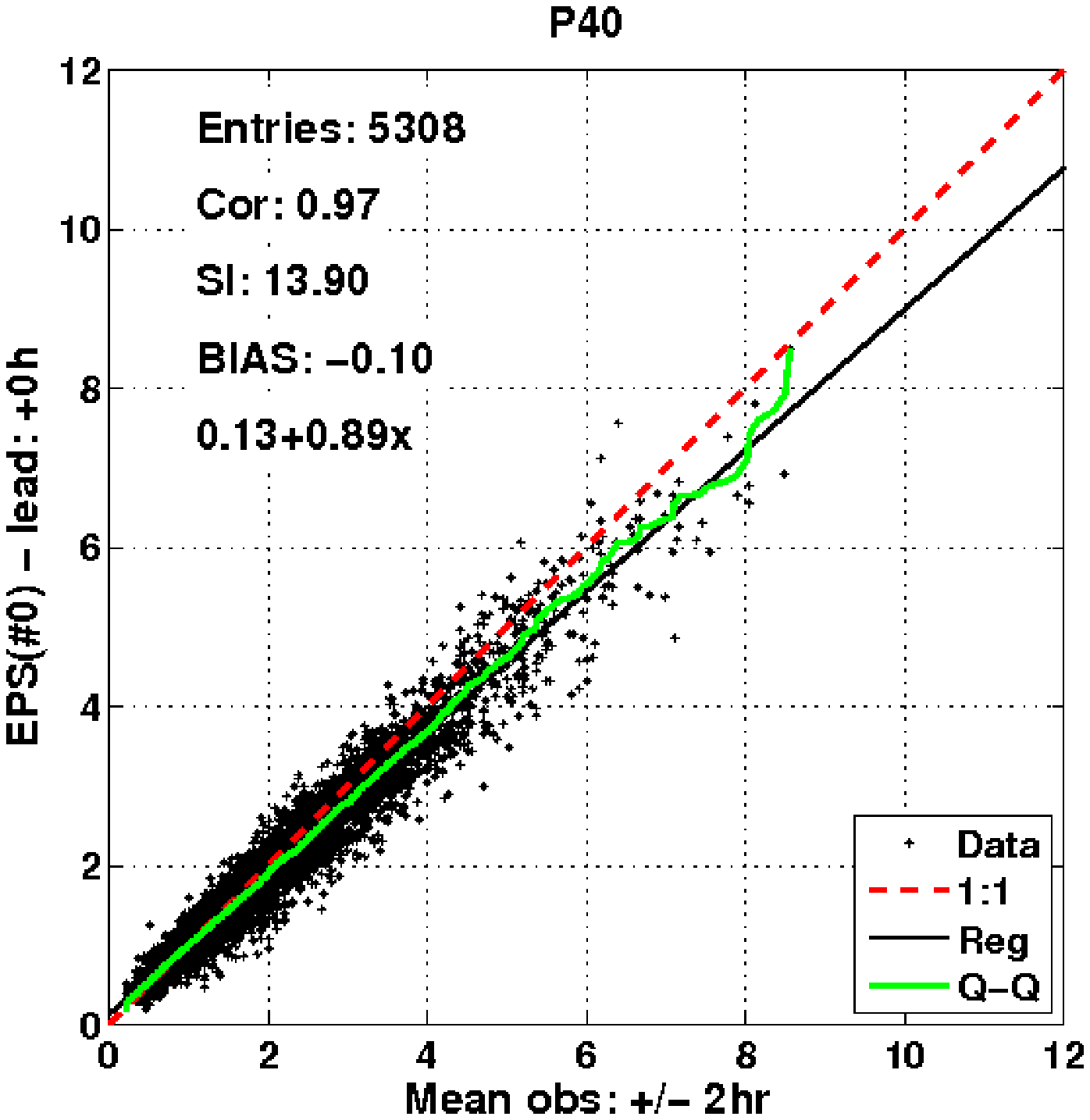}&
(c)\includegraphics[scale=0.45]{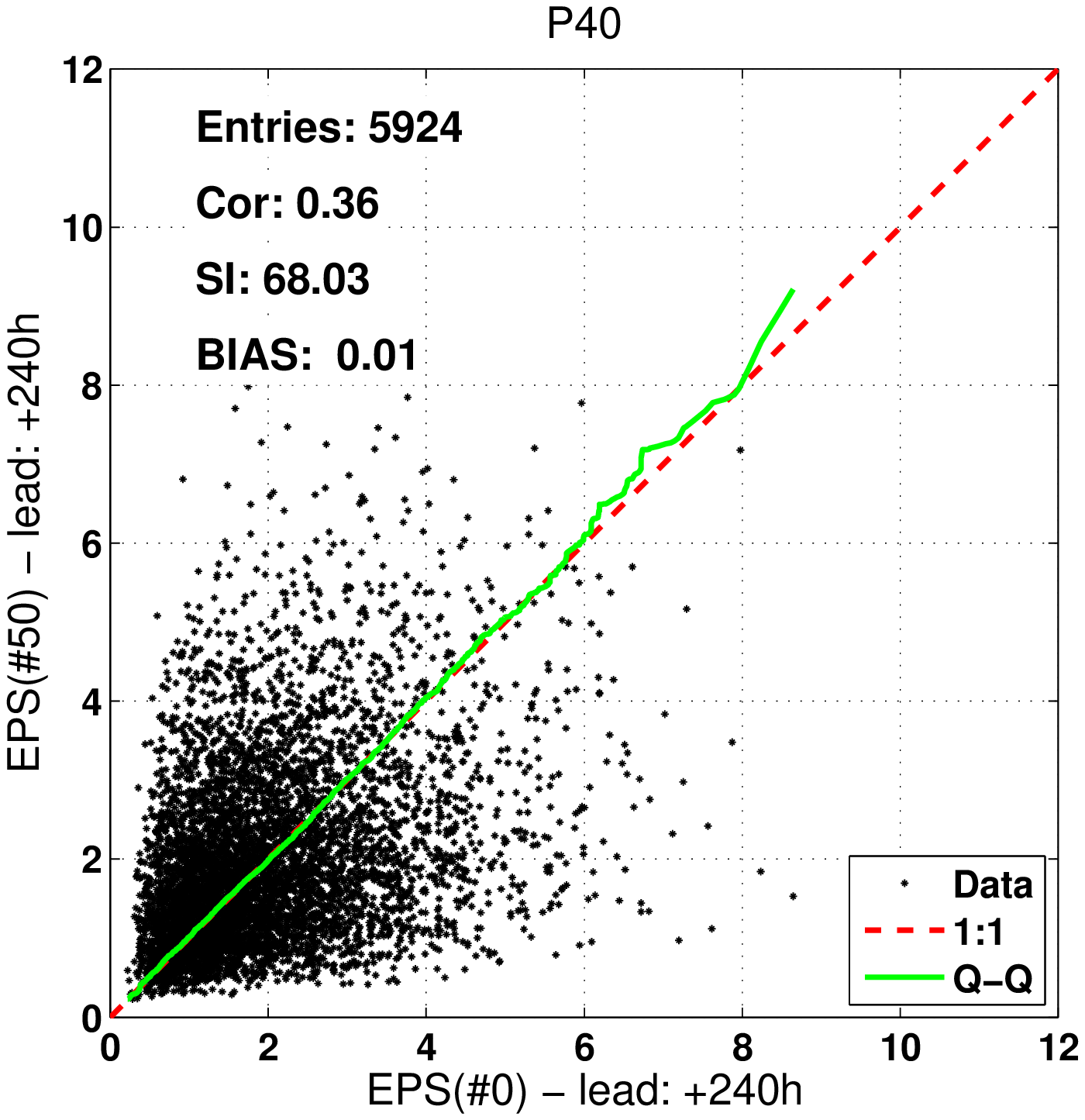}\\
\end{tabular}
\caption{
Panel a: The correlation of all 51 ensemble member forecasts at +240 h with
observations of significant wave height at location P40 (Ekofisk, central North
Sea) over the whole period 1999-2009. The QQ curve is shown in green.  It is
clear that the +240 h climate is quite similar to the observed climate in this
location and better than the wave height distribution found at analysis time
(b).  Panel b: Same as panel (a) for analysis time. The EPS is biased low at
analysis time.  Panel c: The correlation between the +240 h forecasts of two
ensemble members (member 0 represents the unperturbed atmospheric integration)
over the whole period 1999-2009. The QQ-curve is shown in green.  The centered
anomaly correlation relative to the weekly observed wave climate is 0.20. }
\label{fig:corr}
\end{center}
\end{figure}

\begin{figure}
\begin{center}
\begin{tabular}{cc}
(a)\includegraphics[scale=0.45]{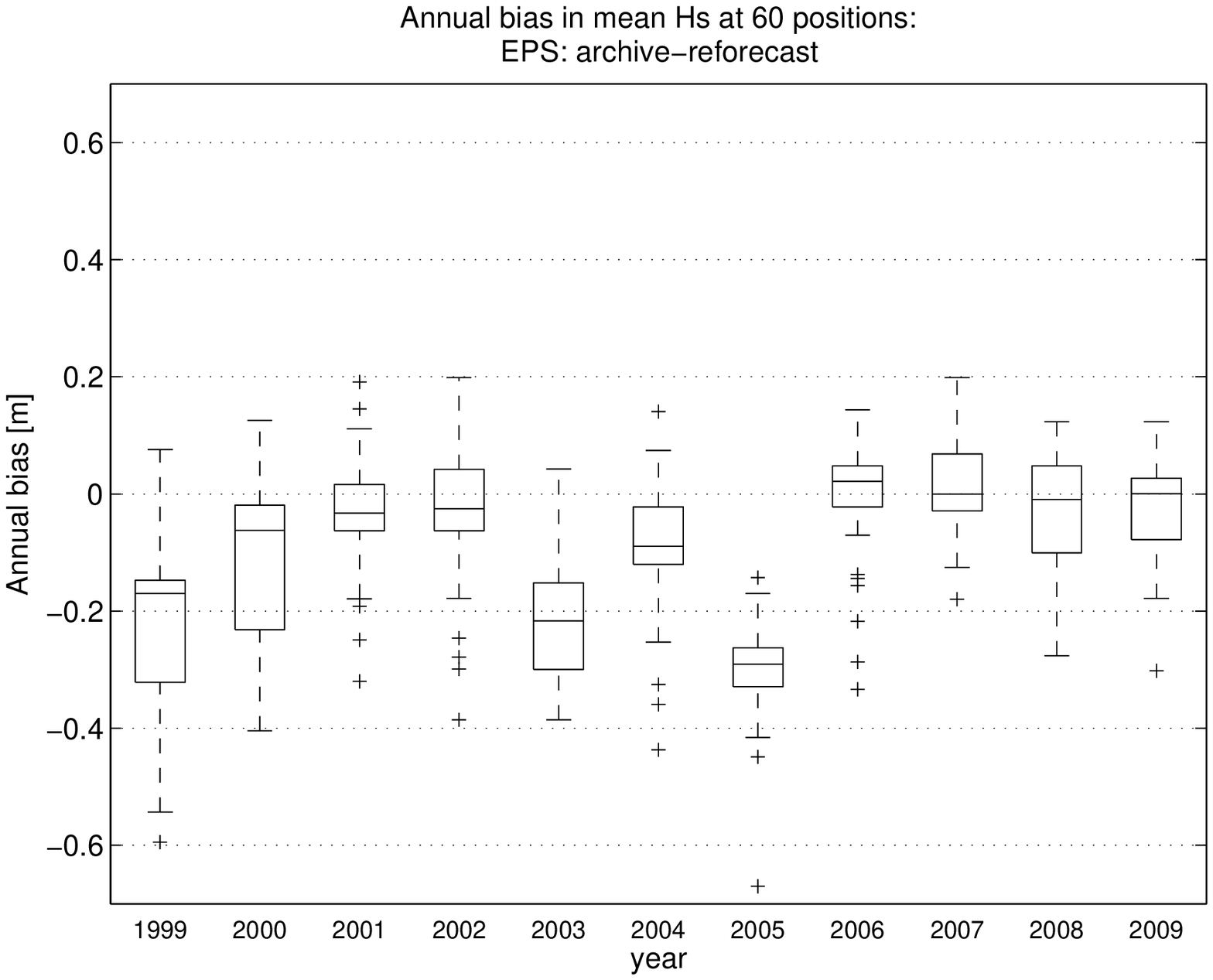}&
(b)\includegraphics[scale=0.45]{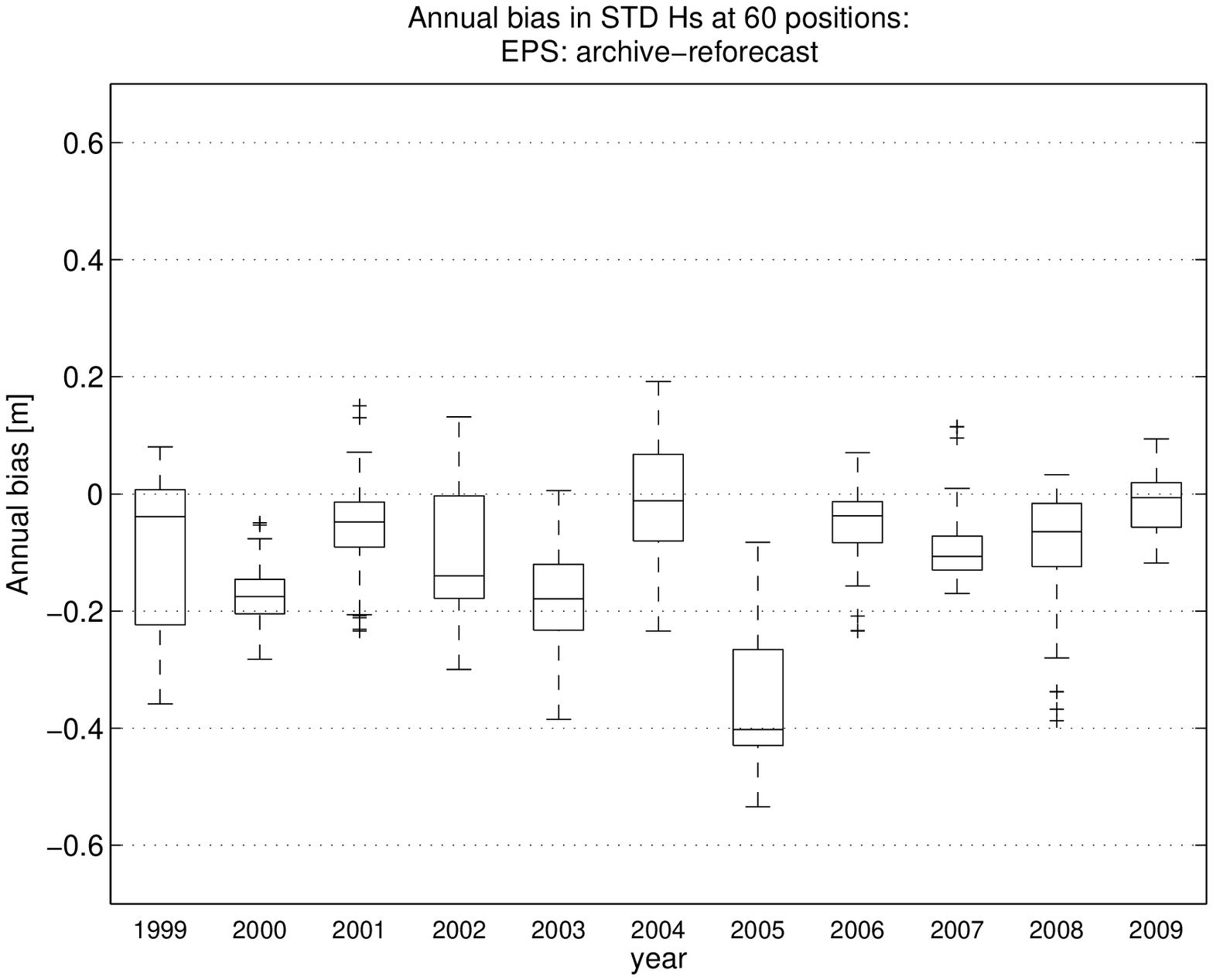}\\
\end{tabular}
\caption{
Panel a: Time series (1999-2009) of the difference in annual mean significant
wave height [m] of EPS forecasts at lead time +240 h and the reforecast at
the same lead time from model cycle Cy36r4 (operational in May 2011) at 60
locations in the north-east Atlantic, the North Sea and the Norwegian Sea.
Panel b: The difference in annual standard deviation [m] of the significant wave
height of EPS240 and the reforecast.  Considerable variations are found from
one year to another, but no significant drift due to model upgrades is seen.}
\label{fig:reforecast}
\end{center}
\end{figure}

\begin{figure}
\begin{center}
\includegraphics[scale=0.6]{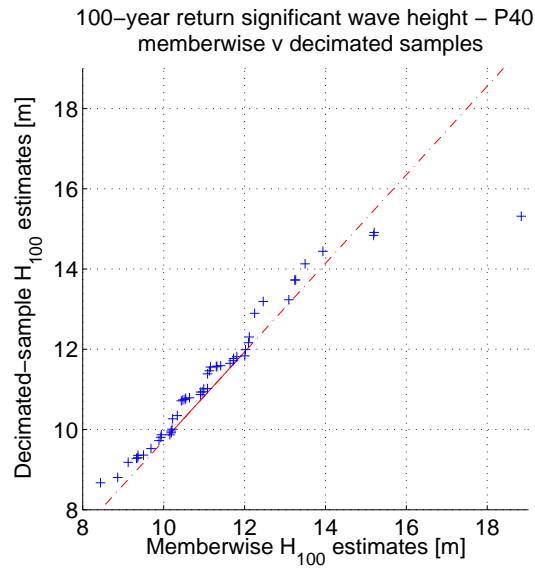}\\
\caption{
Comparison of the quantiles of $H_{100}$ (100-year return values) from 51
ensemble members over the whole period 1999-2009 v the quantiles of $H_{100}$
estimates from 51 decimated samples of all ensemble members taken every 51
forecasts. This decimation gives 51 estimates of same sample size as the
member-wise estimates. The distributions are very similar with common 
averages (11.30~m) and standard deviations of 1.83~m and 1.67~m, respectively.}
\label{fig:h100}
\end{center}
\end{figure}

Gridded estimates of the 100-year return value $H_{100}$ of the significant
wave height were made from EPS240 interpolated to a $1.0^\circ$ grid for the
North Atlantic, the Norwegian Sea and the North Sea using both blocked maxima
(GEV) and threshold exceedances (GP). Note that this is an extraction procedure
and does not reflect the underlying model resolution, which as \Fig{timeline}
shows has increased over the archived period.  Ice-infested locations, i.e.,
locations where the modeled ice concentration ever exceeds 30\%, have been
removed from the analysis.

\begin{figure}
\begin{center}
\begin{tabular}{cc}
(a)\includegraphics[scale=0.45]{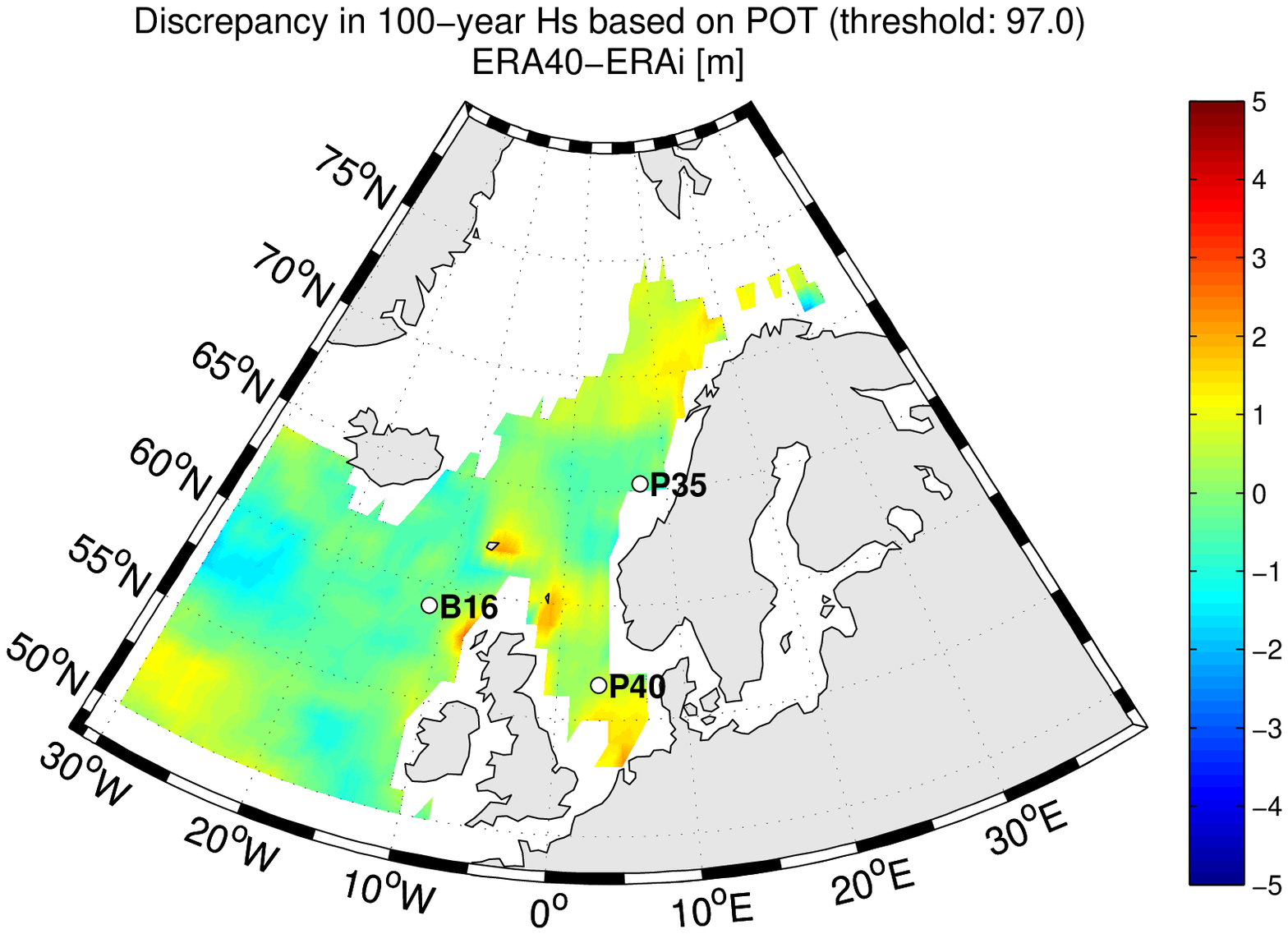}&
(b)\includegraphics[scale=0.45]{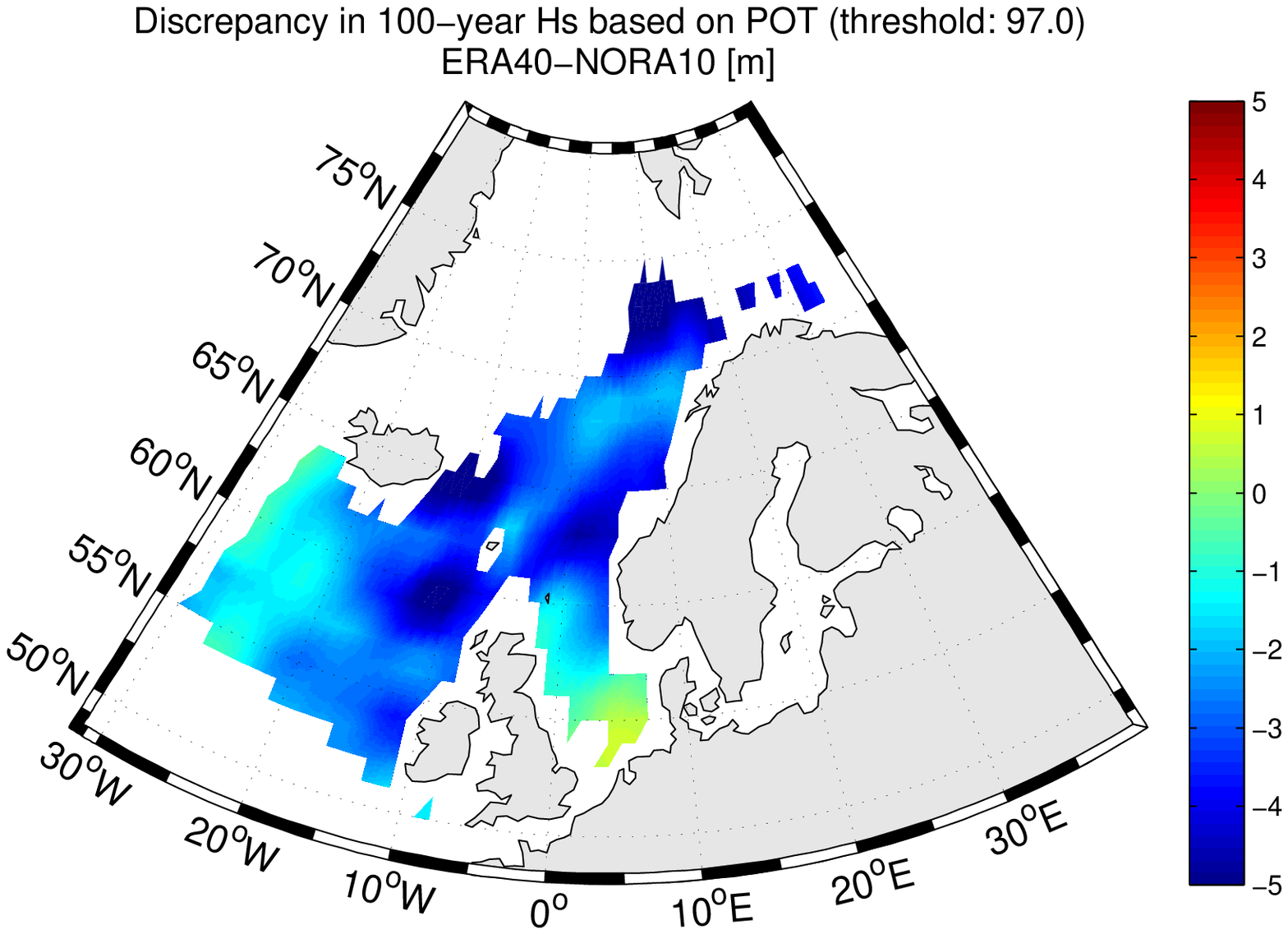}\\
\end{tabular}
\caption{
Panel a: Difference of $H_\mathrm{100}$ of ERA-40 and ERA-I using the GP (POT)
method with a threshold of 97\%. The differences are ${\sim}1\,\mathrm{m}$
in the open ocean, but behind the Faroes and the Shetland isles ERA-40 yields
higher return values. This is to be expected from a coarser reanalysis. The
effect of the deep-water wave physics employed by the ERA-40 WAM model
also yields higher return values in the North Sea.  The model fields are
interpolated to a common grid with $1.0^\circ$ resolution, and all ice-infested
grid points have been excluded.  The three buoy locations are marked as P40
(Ekofisk), B16 (K5) and P35 (Heidrun).  Panel b: Difference of ERA-40 and
NORA10, same threshold as for panel (a). The difference between NORA10 and
ERA-40 is consistently ${\sim}3\, \mathrm{m}$ in the open ocean, but approaches
zero at the open boundary where ERA-40 provided the boundary values.}
\label{fig:era40-erai-nora10}
\end{center}
\end{figure}

We start by looking at the differences we can expect from the existing
reanalyses and hindcasts available to us.  \Fig{era40-erai-nora10} shows the
difference between return values estimated from ERA-40 and ERA-I (panel a) and
similarly between ERA-40 and NORA10 (panel b) using GP (POT) with a threshold
of 97\%. The differences between ERA-40 and ERA-I (panel a) are moderate in
the open ocean, ${\sim}1\,\mathrm{m}$, but the enhanced resolution
of ERA-I is clearly visible on the lee side of the Faroe and Shetland
archipelagos. Also, ERA-40 yields higher estimates in the central North Sea
due to the deep-water physics scheme employed by the ERA-40 WAM model. Panel
(b) shows the difference between ERA-40 and NORA10. As reported earlier by
\cite{aar12}, NORA10 consistently estimates higher return values than ERA-40
(3-4 m higher in the northern North Atlantic), except along the
open boundary to the south and west where the model is forced with ERA-40
boundary values. Note also that the central North Sea exhibits the smallest
difference, since the ERA-40 deep-water physics yields higher values here.
\cite{rei11} looked at the upper percentiles of the NORA10 wave height
distribution in a number of locations in the North Sea and the Norwegian
Sea and found good agreement with buoy measurements.  It is clear that the
differences seen around the Faroes and Shetland in \Fig{era40-erai-nora10}
are due to resolution issues partly solved by ERA-I and more fully resolved by
NORA10. In the open ocean ERA-40 and ERA-I seem to be in agreement while
NORA10 estimates significantly higher return values (on the order of 3-4 m in
the northern North Atlantic, consistent with what is found by \citealt{aar12}).

\begin{figure}
\begin{center}
\begin{tabular}{cc}
(a)\includegraphics[scale=0.45]{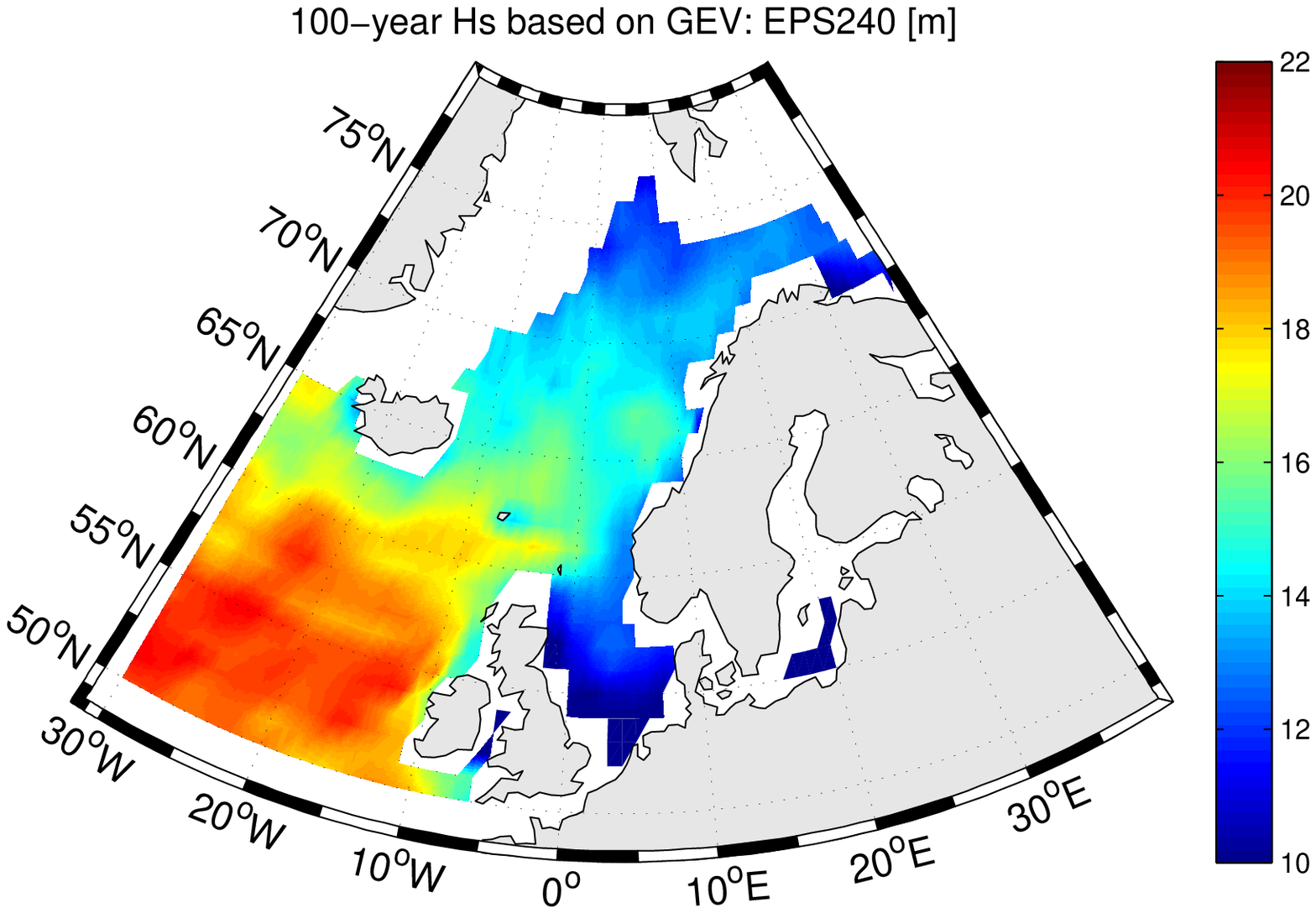}&
(b)\includegraphics[scale=0.45]{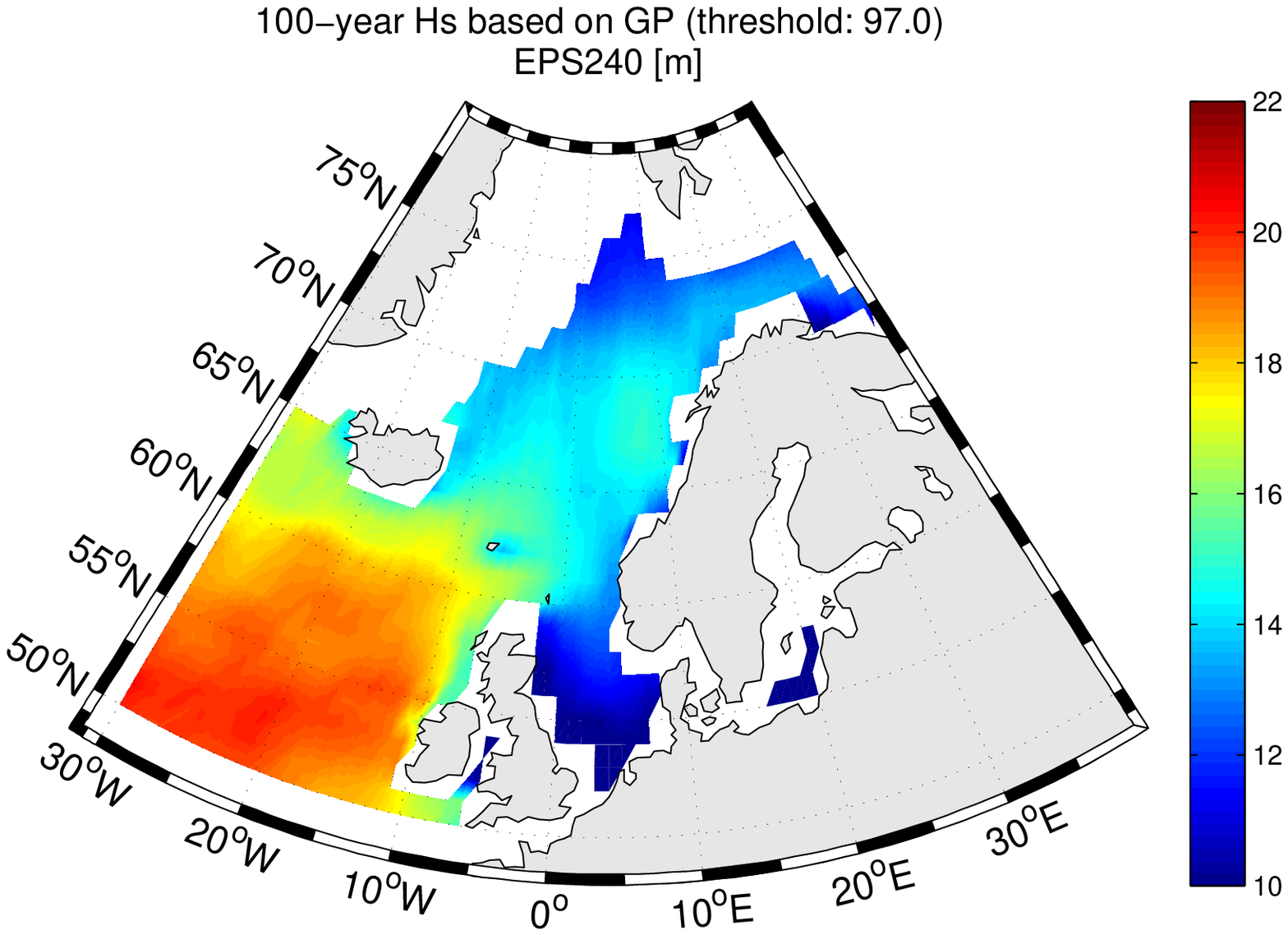}\\
\end{tabular}
\caption{
Panel a: Gridded estimates of $H_\mathrm{100}$ from EPS240 using GEV with
blocked maxima from individual ensemble members. The grid is $1.0^\circ$,
and all ice-infested grid points have been excluded.  Panel b: Same as panel
(a) but for GP with a threshold of 97\% of the data.
}
\label{fig:H100_eps240}
\end{center}
\end{figure}

\Fig{H100_eps240} presents the return values for EPS240 using GEV (panel a)
and GP (panel b). The differences are very small. The blocking for GEV was
performed per ensemble member to minimize the effect of the varying resolution
and the numerous model upgrades introduced throughout the archived period
(cf \Fig{timeline}). The GP threshold was set at $P_{97}$, but varying this
threshold did not affect the return values considerably (results not shown).
\cite{caires05b} and \cite{sterl05} find that ERA-40 when calibrated
against buoy data yields return values on the order of 24~m in the
northern North Atlantic; higher but much closer to our findings than the
uncorrected ERA-40 return values.

\begin{figure}
\begin{center}
\begin{tabular}{cc}
(a)\includegraphics[scale=0.45]{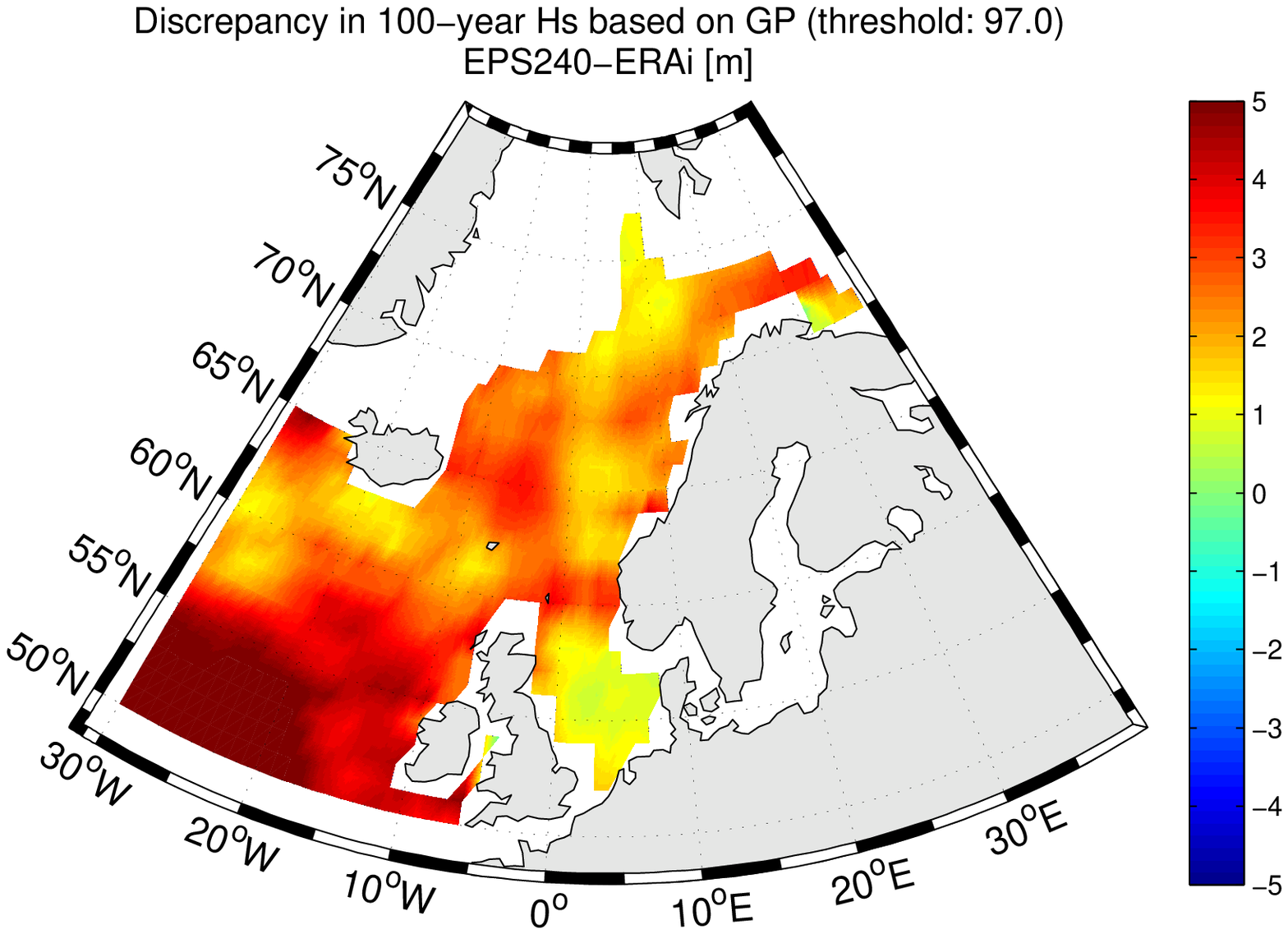}&
(b)\includegraphics[scale=0.45]{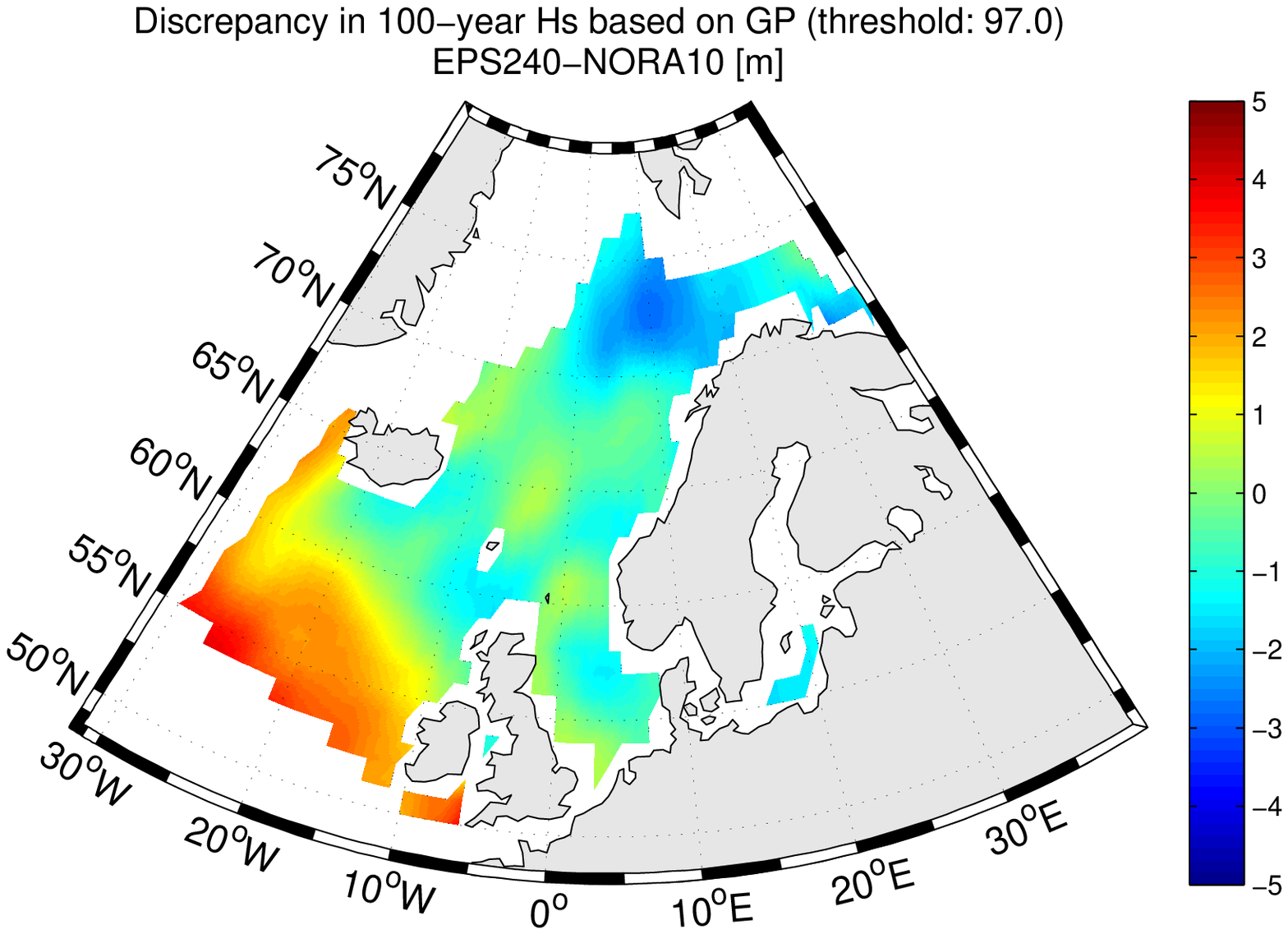}\\
\end{tabular}
\caption{
Panel a: 
Difference between $H_{100}$ estimates of EPS240 and ERA-I using
the GP distribution with a threshold of 97\% (positive means EPS240 is higher). EPS240 consistently
predicts higher return values throughout the domain. The differences
are ${\sim}3\,\mathrm{m}$ in the open ocean, with the North Atlantic
approaching a difference of 5 m. The differences are smallest in the central
North Sea.  The grid is $1.0^\circ$, and all ice-infested grid points have
been excluded.
Panel b:
Difference between EPS240 and NORA10, same GP threshold as for panel
(a). The difference between NORA10 and EPS240 is generally smaller
than what was found in panel (a) for ERA-I, but significant geographical
differences exist. Near the south-western boundary NORA10 is influenced by
ERA-40, and behind the Faroe and Shetland archipelagos the resolution of
EPS240 is too coarse to provide a meaningful comparison.
}
\label{fig:eps240-erai-nora10}
\end{center}
\end{figure}

\Fig{eps240-erai-nora10} presents the difference between the GP 100-year
return values found for EPS240 and those found for ERA-I (left panel) and
NORA10 (right panel) using GP with a $P_{97}$ threshold. It is clear that
EPS240 predicts considerably higher return values than what is found for
ERA-I. The differences approach 5~m in the North Atlantic, and throughout
the Norwegian Sea we see differences on the order of 2-3~m. The differences
between NORA10 and EPS240 are smaller (panel b), but here we see significant
differences throughout the domain that must be considered separately. First,
the influence of ERA-40 on NORA10 is visible along the open boundary in
the south-west, so the boundary zone should not be taken into account in
this analysis. Second, EPS240 does not properly resolve the Faroes and
the Shetland archipelagos. The middle North Sea is biased low, which is
probably also a resolution issue. Away from these areas we see that the
agreement is generally good, except in the northern part of the Norwegian
Sea, where EPS240 is up to 3~m lower than the NORA10 estimates.  
\cite{aar12} discusses the large impact of an individual
storm event on the NORA10 estimates in this area and we note that the 
confidence intervals are exceptionally wide here (14-22~m, see \citealt{aar12},
Fig 3).

\subsection{Bootstrapping confidence intervals}
The GEV shape parameter $\xi$ in \Eq{gev} and its counterpart for GPD in
\Eq{gp} determine the width of confidence intervals \citep{hosking84,col01}.
The significant wave height from the NORA10 hindcast has been shown by
\cite{aar12} to exhibit a wide range of extreme value shape parameters
within the Norwegian Sea and the adjacent seas with correspondingly varied
confidence intervals.

We have estimated confidence intervals for EPS240 and ERA-I using a
bootstrapping technique similar to that employed by \cite{aar12}.  For ERA-I
which represents a traditional time series we have made 100 random draws with
replacement from the POT data (see Fig~\ref{fig:eps240-erai-nora10}). In
the case of EPS240, we have similarly made random draws from the tail of
the dataset exceeding the 97 percentile (note that this is technically not
\emph{peaks}-over-threshold since the EPS240 data are considered independent.

\begin{figure}
\begin{center}
\begin{tabular}{cc}
(a)\includegraphics[scale=0.45]{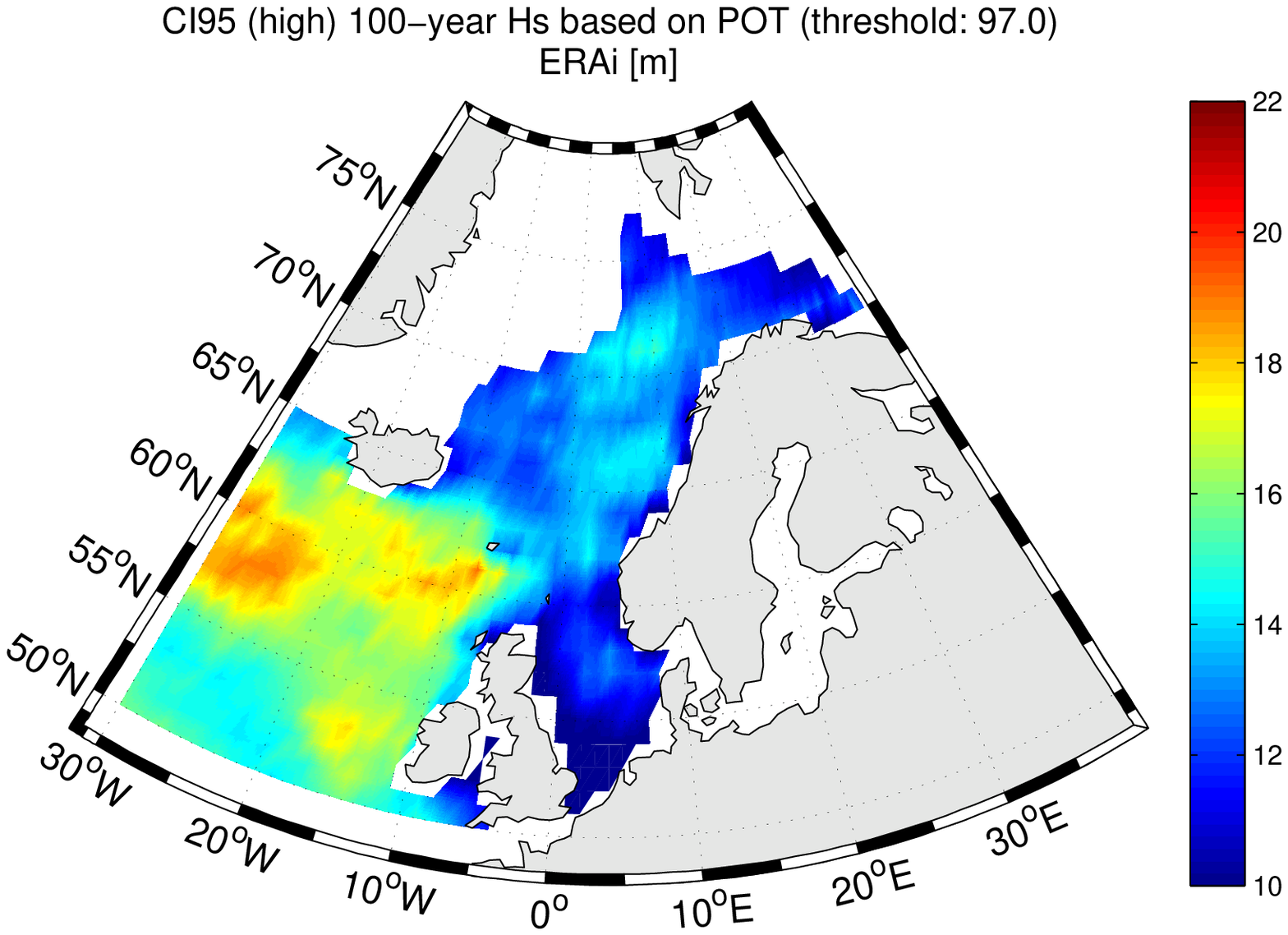}&
(b)\includegraphics[scale=0.45]{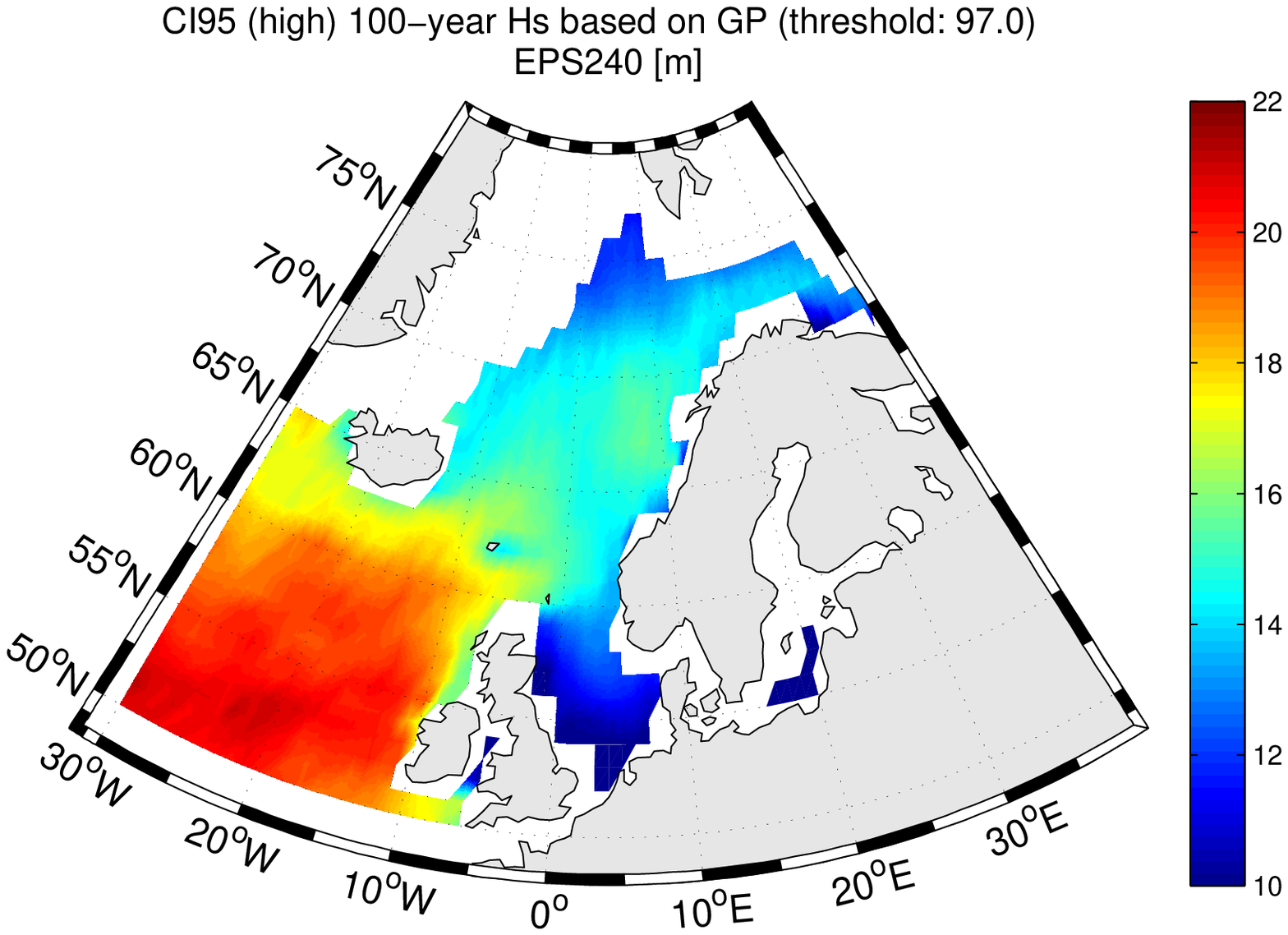}\\
\\
(c)\includegraphics[scale=0.45]{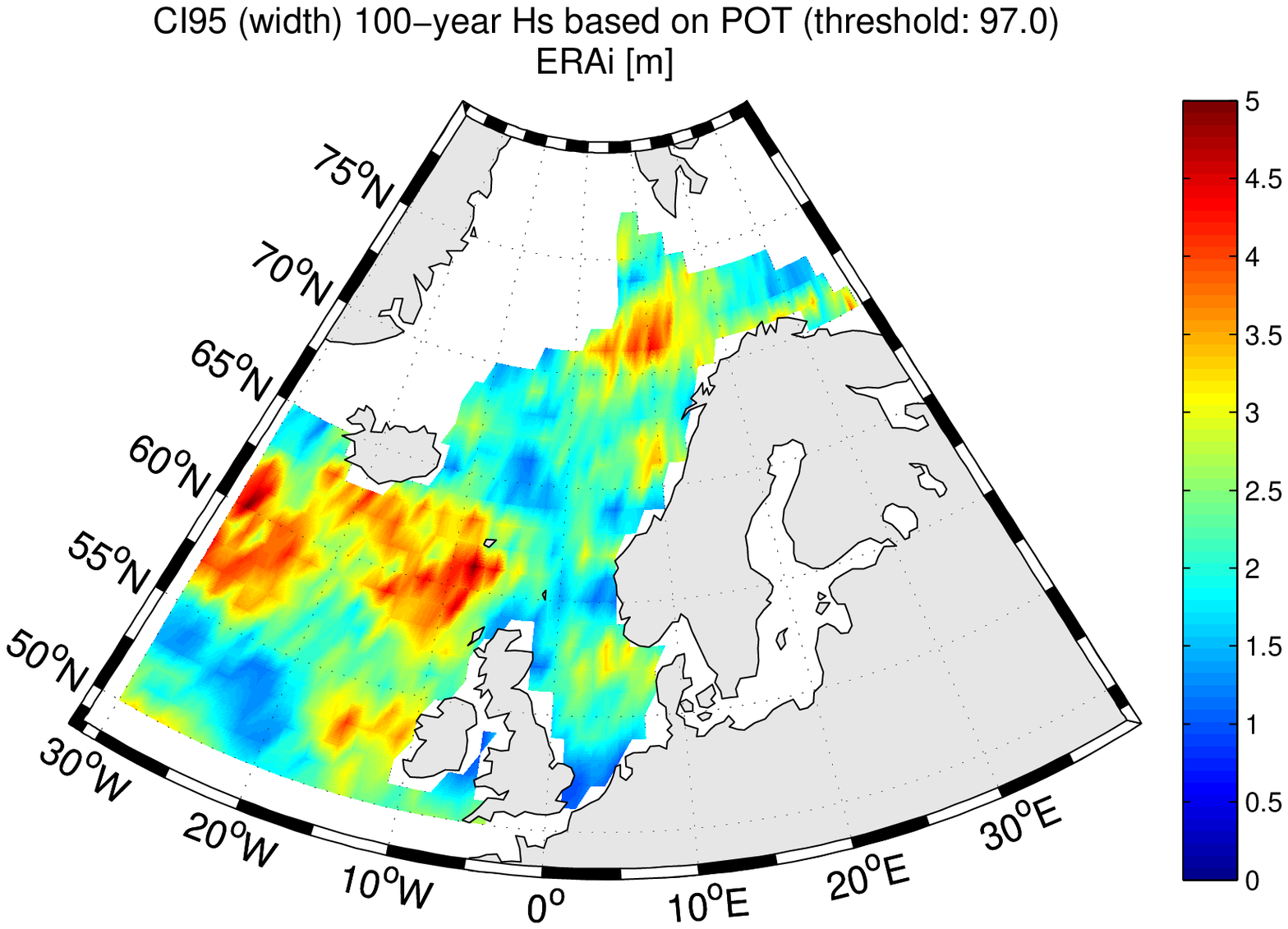}&
(d)\includegraphics[scale=0.45]{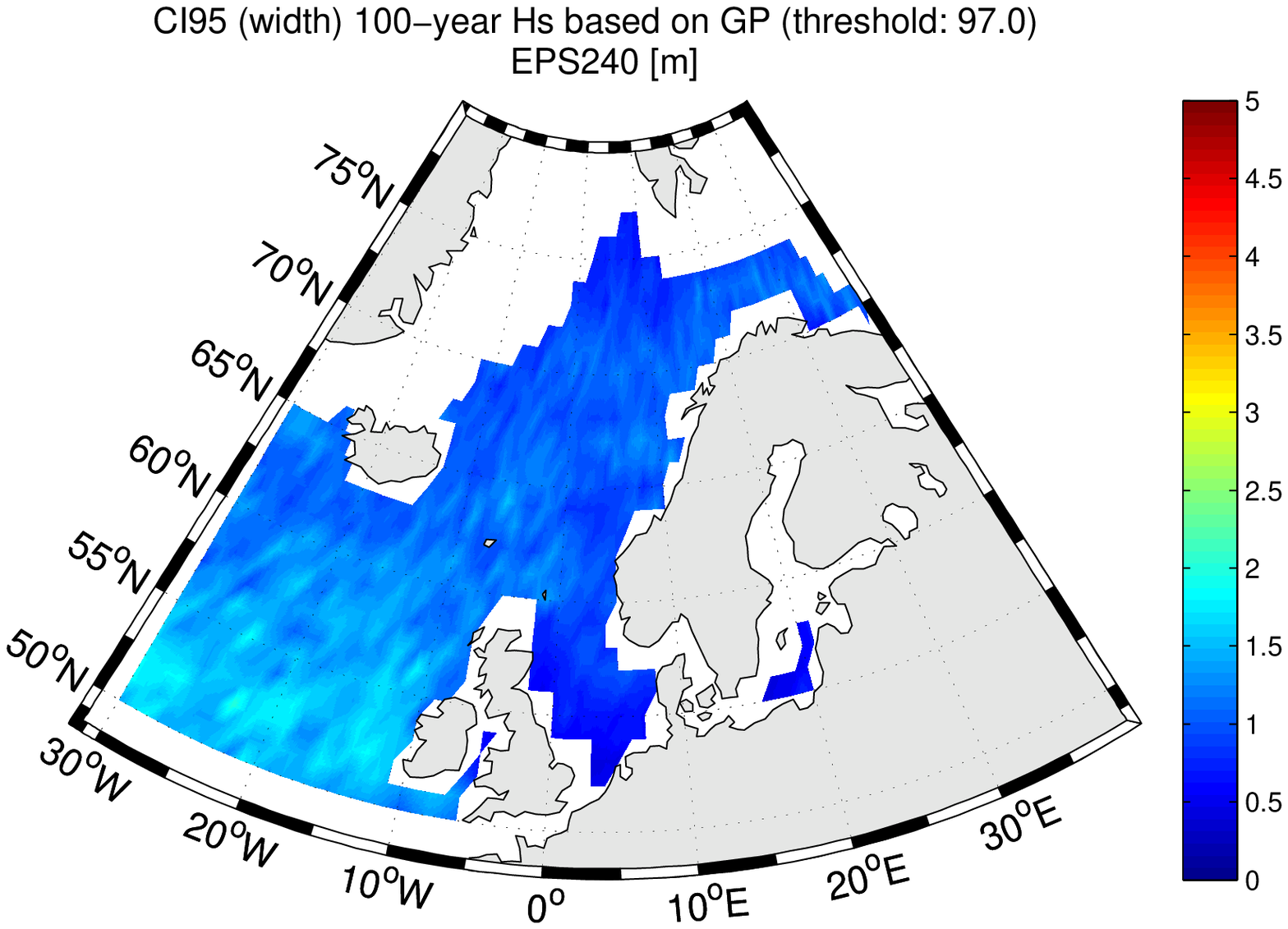}\\
\end{tabular}
\caption{
Panel a: Upper limit of 95\% confidence interval for ERA-I $H_{100}$ based
on 100 bootstraps of the POT exceeding the 97 percentile.
Panel b: Upper limit of 95\% confidence interval for EPS240 $H_{100}$ based
on 100 bootstraps of the data exceeding the 97 percentile.
Panel c: Width of 95\% confidence interval for ERA-I.  The relative width
reaches 30\% of the return values in parts of the north-east Atlantic. The
geographic variability is pronounced, largely due to influence from individual
storm events.  Panel d: Width of 95\% confidence interval for EPS240.
The relative width varies from 5\% in sheltered areas to 10\% in the open
ocean. The geographic variability is very low.}
\label{fig:ci95}
\end{center}
\end{figure}

The upper limits of the confidence intervals found for ERA-I and
EPS240 are shown in \Fig{ci95}, panels~(a) and~(b).  The differences are
pronounced. First, the confidence interval is much tighter for EPS240 (panel~d), ranging from less than 1~m in the sheltered parts of the North Sea to
approximately 2~m in the open ocean (relative width 5-10\% of the return
values). ERA-I (panels~a and~c) has confidence intervals up to 5~m (relative
width 30\% of the return values) in the north-east Atlantic. Second,
the spatial variability of the confidence intervals is very low for EPS240,
while the ERA-I intervals vary substantially throughout the domain due to
sensitivity to individual storm events.

It is important to stress that even though the confidence intervals become
much tighter with a larger data set, the bootstrapping method does not
account for model bias. The bias must be assessed by comparing the observed
and modeled wave height distributions, see \Sec{method}(\ref{sec:criteria}).
We discuss the impact of model bias further in \Sec{discussion}.

\section{Strengths and limitations to the method}
\label{sec:discussion}
Estimating return values from ensembles at advanced lead times is a new
technique, and the assumptions underlying the method have been outlined
in \Sec{method}. Here we discuss some of the perceived weaknesses of the
method in general and how applicable the method appears to be for significant
wave height.  The main caveats to be aware of when using the technique on
archived EPS forecasts in general are:
\begin{enumerate}
 \item Spurious trends caused by model upgrades
 \item Upper-percentile biases
 \item Conversion to an equivalent time period
 \item Correlations within the ensemble
 \item Return value estimates in a changing climate
\end{enumerate}
Although we do not find evidence of any spurious trend in the mean and the
variance of the significant wave height at +240 h (see \Fig{reforecast}),
we are aware that the IFS model updates over the past decade have led to
an apparent increase in the 10-m wind speed at analysis time. S.~Abdalla
(pers comm) quantified this effect at analysis time to be about $29\,
\mathrm{cm}\,\mathrm{s}^{-1}$, i.e., the earlier analyses were biased low.
The wave height will be somewhat affected by this, but it is thought to
have a small effect on the extremes of waves found at advanced lead times,
especially since some of the removed bias stems from changes to the data
assimilation and will fade as the model integration becomes dynamically
balanced at advanced lead times. The effect is also evident from inspection
of \Fig{corr} where the wave height at analysis time (panel b) is seen to
be biased low.  Since this bias disappears for EPS240 (panel a), we believe
that the model updates have had only a modest impact on the wave climatology
at advanced lead times.

\begin{figure}
\begin{center}
\begin{tabular}{cc}
(a)\includegraphics[scale=0.5]{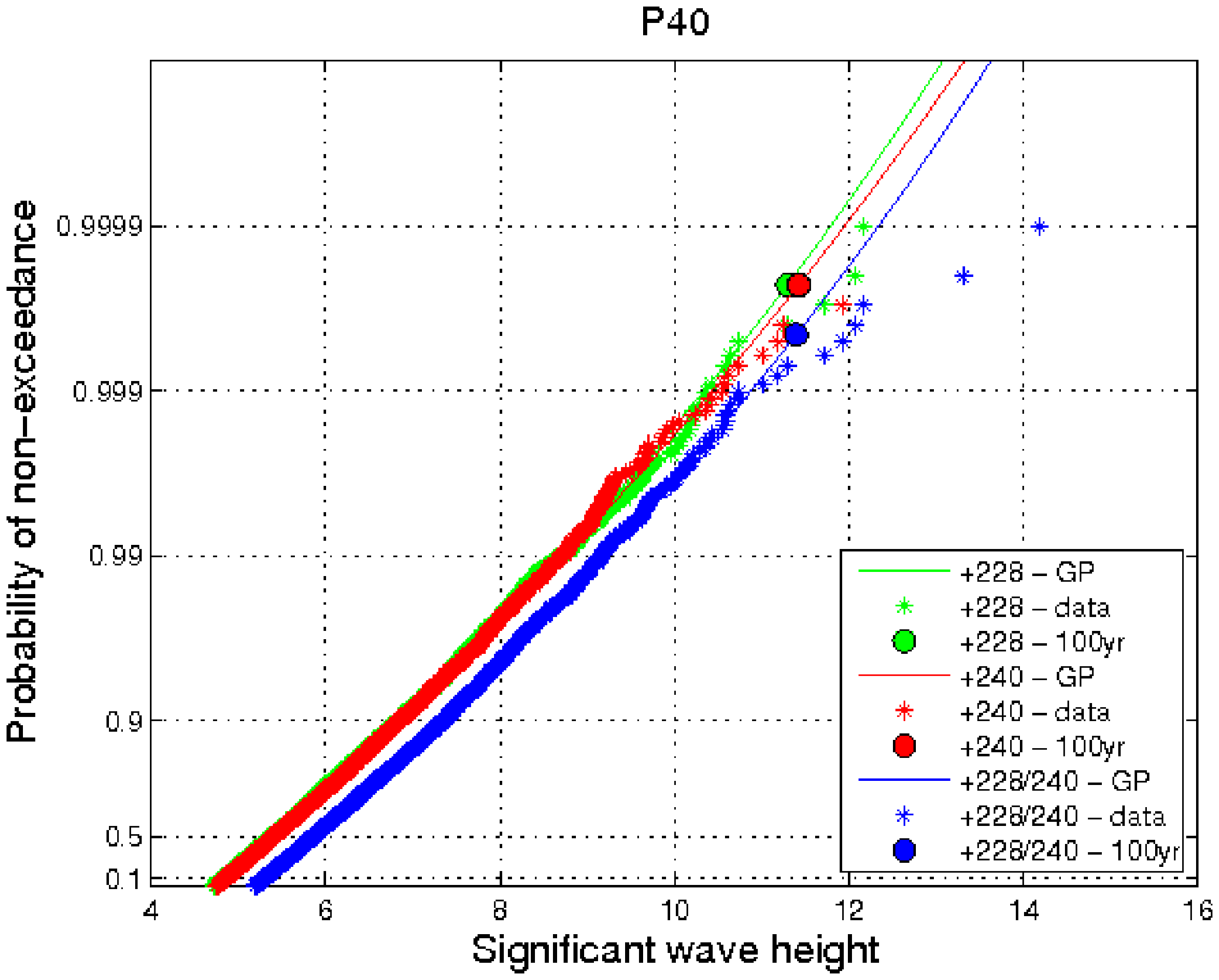}&
(b)\includegraphics[scale=0.5]{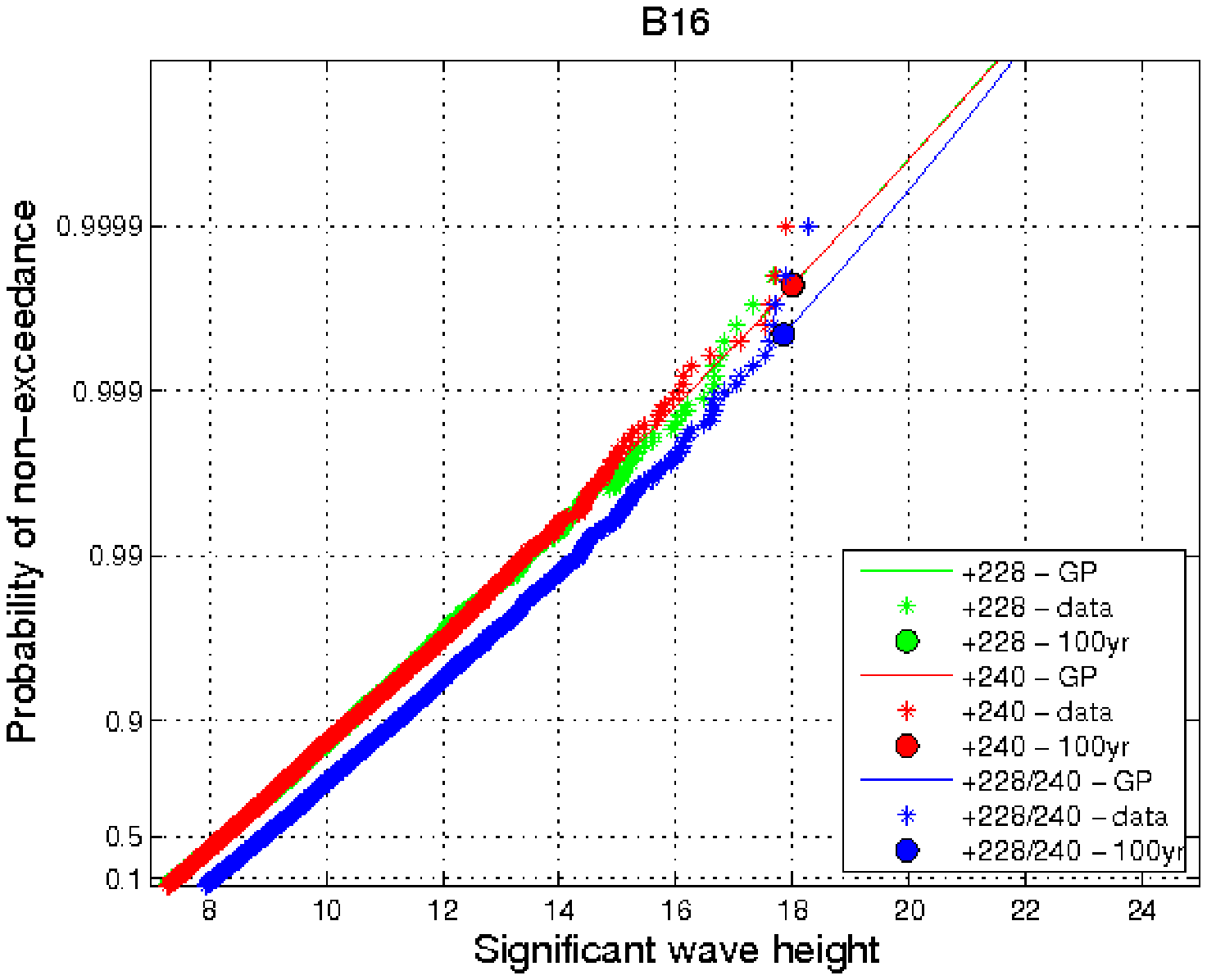}\\
\\
(c)\includegraphics[scale=0.5]{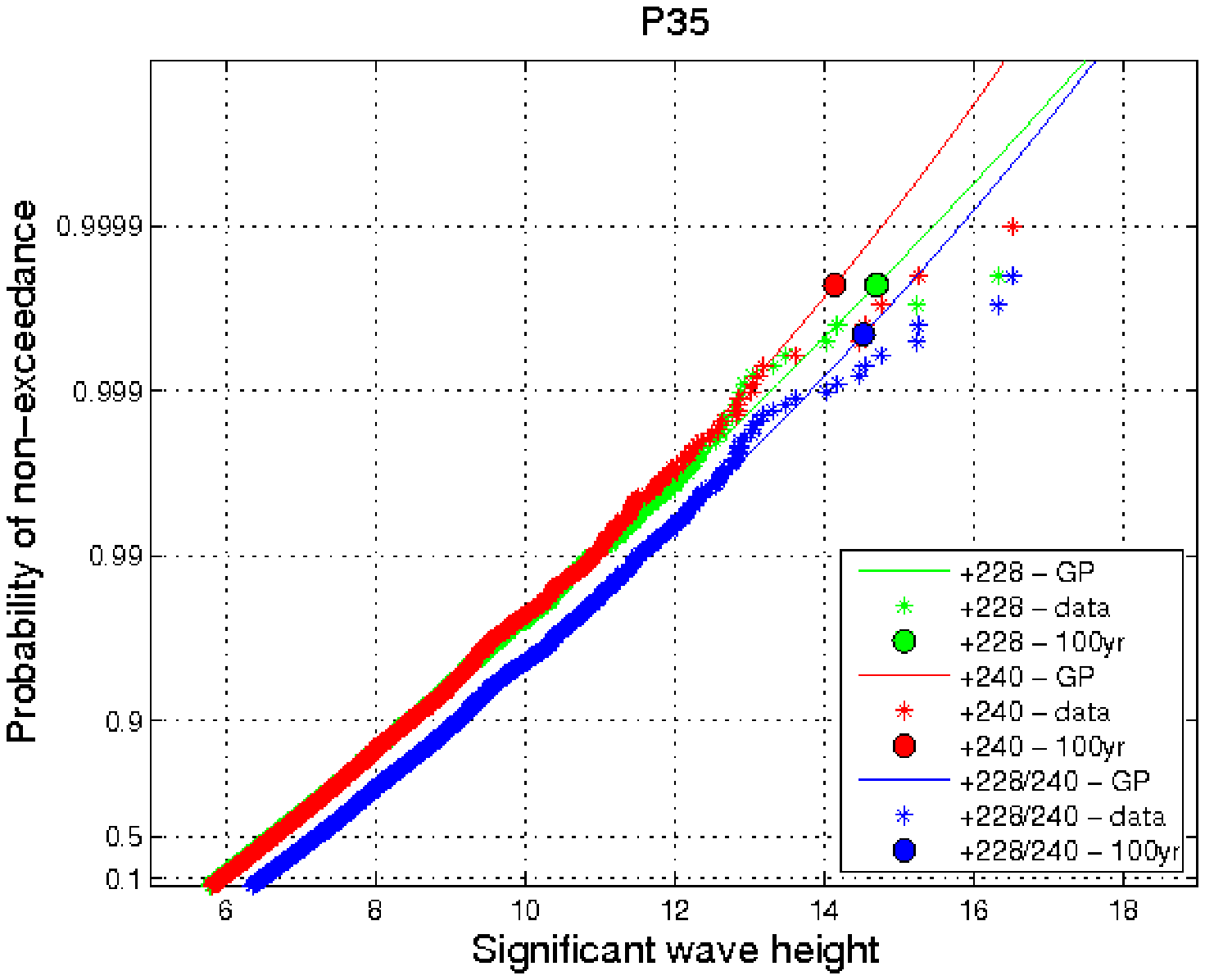}\\
\end{tabular}
\caption{
Panel a: Probability of non-exceedance for the EPS forecasts in location
P40 (Ekofisk, central North Sea). The GPD estimates are shown as curves
(with 100-year values marked as circles) and are based on a 97\% threshold
while the individual forecast values are shown as asterisks.  Green and
red indicate EPS228 and EPS240, respectively. Blue is a combined estimate
for EPS228 and EPS240 where the maximum of each pair of EPS228 and EPS240
forcast is chosen.  This is done because the two forecasts are separated by
only 12 h and strongly correlated.  The combined data set thus represents
the equivalent of 452 years of data since EPS228 and EPS240 each represents
the equivalent of 226 years.  The combined data set lies below the EPS228 and
EPS240 on the vertical axis since it has a lower probability of non-exceedance
due to being twice the size of EPS228 and EPS240.  Panel b: Same as panel (a)
but for location B16 in the eastern North Atlantic (note that the upper three
values of EPS228 are masked by EPS240 as the values are almost identical.
Panel c: Same as panel (a) but for location P35 in the eastern Norwegian Sea
(Heidrun).  It is evident from all three panels that the combined 100-year
estimate is bracketed by the estimates from the two individual estimates, as 
expected from a larger dataset.}
\label{fig:exceedance}
\end{center}
\end{figure}

We have investigated the robustness of the return value estimates by
also looking at EPS228. We select the maximum from each pair of EPS228
and EPS240 since the +228 and +240-h forecasts are strongly correlated
(see \Fig{exceedance}). This combined data set is now assumed equivalent
to $2\times 226$ years. The combined 100-year GP return value estimates
(indicated by blue circles) fall between the 100-year return values from
the the two data sets (EPS228, green circles, and EPS240, red circles),
which is what we expect when going to larger data sets. This suggests that
even larger data sets may be built by selecting maxima from longer forecast
sequences. However, care must be taken to avoid getting too close to the
beginning of the forecast where the ensemble members are correlated.

\begin{figure}
\begin{center}
\begin{tabular}{ccc}
&(a)\includegraphics[scale=0.5]{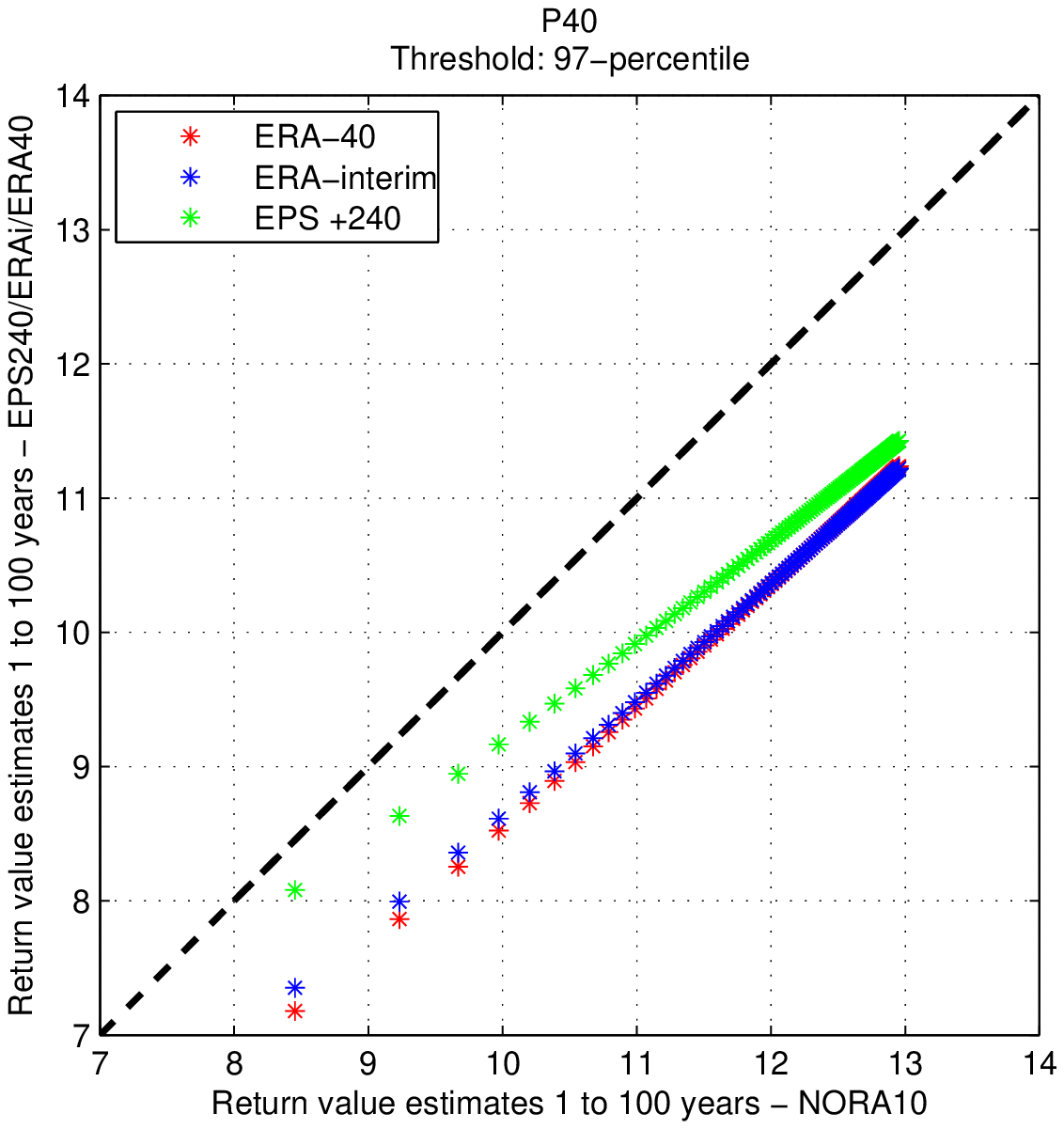}&
(b)\includegraphics[scale=0.5]{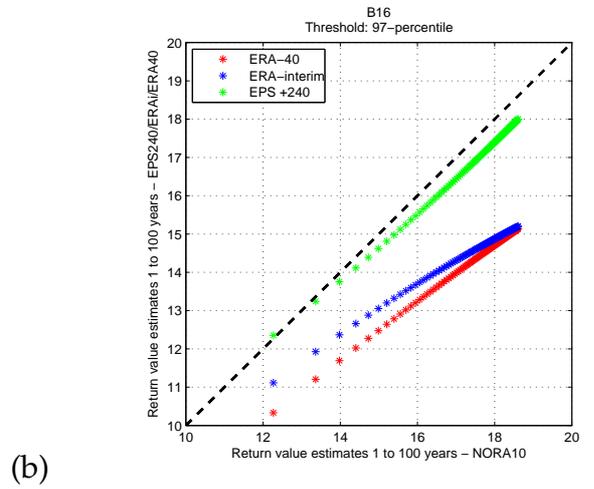}\\
\\
&(c)\includegraphics[scale=0.5]{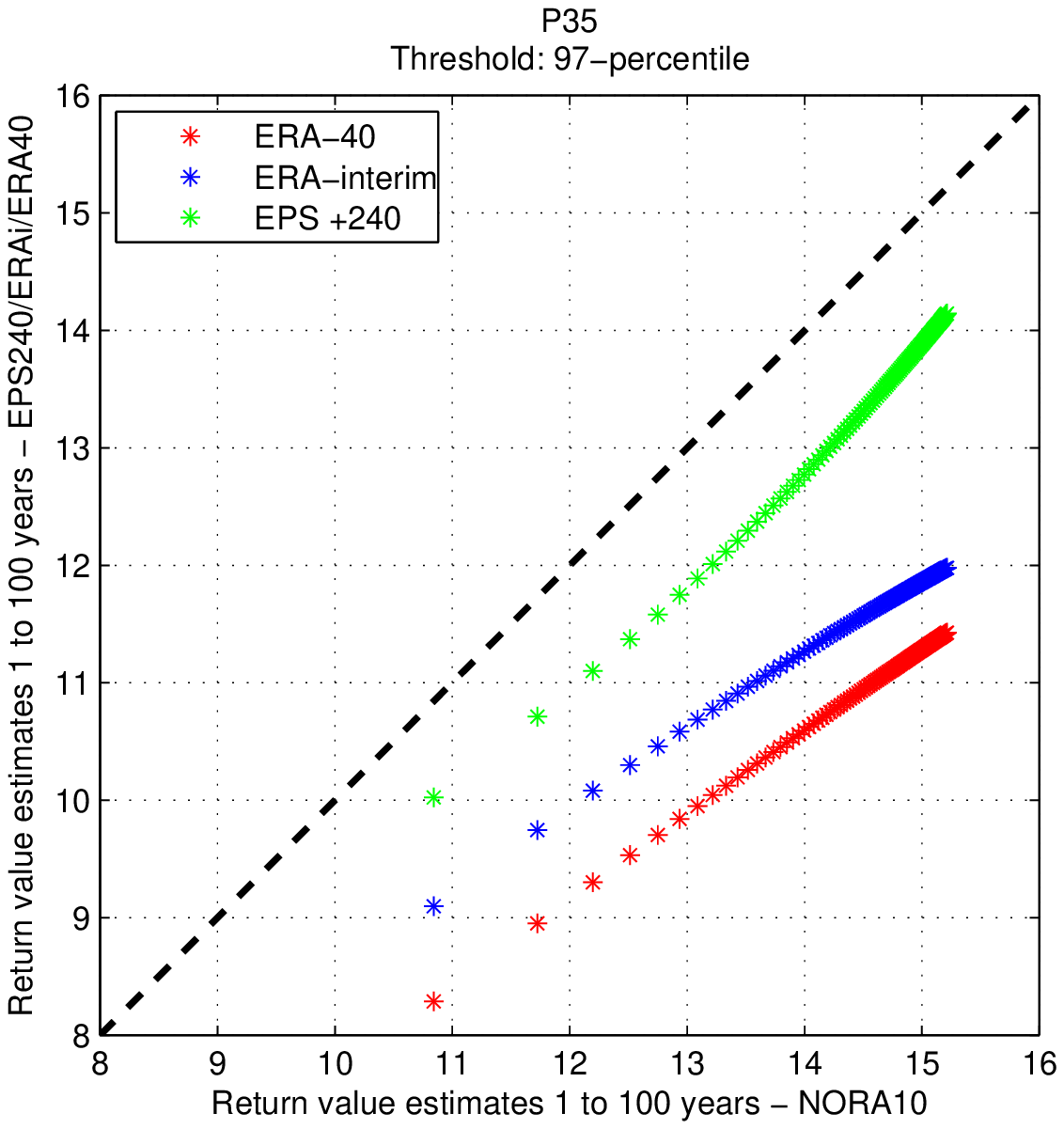}\\
\end{tabular}
\caption{
Panel a. ``Comet plot'' comparison of $H_1,H_2,\ldots,H_{100}$ return values
in the central North Sea (P40, Ekofisk) for NORA10 (first axis) against ERA-40
(red asterisks), ERA-I (blue) and EPS240 (green). All return estimates were made
from the GP distribution with a 97\% threshold.  Panel b. Same as panel (a)
but for location B16 in the eastern North Atlantic.  Panel c. Same as panel
(a) but for location P35 in the eastern Norwegian Sea (Heidrun).  ERA-40 and
ERA-I estimates are significantly lower than NORA10 in all three locations.
EPS240 shows good correspondence in open-ocean conditions over the whole range
up to $H_{100}$ (panels b and c), while in the North Sea the upper return
values are substantially lower and closer to the ERA-40 and ERA-I estimates.}
\label{fig:comet} 
\end{center} 
\end{figure}

We have also looked at the possible sources of bias to the extremes from
EPS forecasts. Such a bias can not be estimated from a bootstrap procedure.
Instead we have compared the return values of the ERA-40, ERA-I and EPS240
against NORA10 which has been shown to represent the upper percentiles well
\citep{rei11,aar12}.  Biases can enter a model data set in two distinct
ways. The first is through poor representation of physical processes. For
example (as pointed out by \citealt{aar12}), ERA-40 applies deep-water physics
in areas where this is questionable (such as the southern North Sea). If
a bias is due to poor model physics or poor model resolution then the bias
should also be different from one region to another.  The second way biases
can enter a data set is through poor spatial and temporal sampling of the
model fields. For example, both ERA-40 and ERA-I are archived with six-hourly
resolution and will consequently miss some modeled storm maxima, even if a
coarse model as discussed in \Sec{data}(\ref{sec:obs}) is slowly varying and
representative of significant wave height averaged over 6~h. The data sets
are also typically interpolated in space, leading to further reduction of
extremes. This means that the return values from EPS and coarse resolution
reanalyses such as ERA-40 and ERA-I should be interpreted as return values of
the six-hourly average sea state, as discussed in \Sec{data}(\ref{sec:obs}),
and will thus generally be lower than those found from a high-resolution
hindcast such as NORA10 where the model values represent shorter time
intervals.  \Fig{comet} shows how the return values ranging from $H_1,
\ldots, H_{100}$ line up. It is clear that ERA-40 and ERA-I due to their
negative bias give significantly lower return values than NORA10, especially
in the open-ocean conditions in the North Atlantic (panel b).  EPS240 on the
other hand matches the return values better, and the shorter return periods
also line up well here. For Ekofisk (panel a) in the central North Sea the
situation is different. Here the shorter return periods match well, but due
to the relatively coarse resolution of EPS240 $H_{100}$ seems to converge
to approximately the same value as ERA-40 and ERA-I (11.3~m). NORA10 yields
100-year return values closer to 13 m for this location. It seems likely
that the EPS240 return values are biased low in enclosed seas and should
consequently be used with some care in such areas. The EPS240 return values
are close to NORA10 estimates in the open ocean and we conclude that the
bias from spatial and temporal interpolation is of less importance, since
otherwise the return values should be depressed everywhere.

Lack of forecast skill at advanced lead times is an important requirement since
the ensemble members must be assumed uncorrelated to be considered independent
draws from the model climate. We have shown that the weak correlations in the
mean are not present in the tail of the distribution in the case of significant
wave height (see \Sec{method}). However, it seems likely that the method is not
equally applicable to the investigation of the extremal behavior of parameters
representing large-scale features, e.g. the North Atlantic Oscillation (NAO)
index \citep{hur95}, or long-term (seasonal, say) averages. Here we do expect
the ensemble forecast system to retain skill at advanced lead times, and
indeed forecast skill in reproducing large-scale features is the rationale
behind seasonal forecast systems \citep{stockdale98,stockdale11}, where the
lead time typically goes to six months \citep{brink05}. We therefore find
it prudent to advice against employing the method on large-scale spatial
averages or long-term temporal averages.  It is also clear that the forecasts
only differ from the initial conditions by as much as 240-h integrations allow
and will still be influenced by the slow components of the earth system, like
the Arctic ice cover.  This means that for parameters influenced by climate
change or where quasi-cyclical phenomena with long-periodic components such
as the El Ni\~no, Southern Oscillation (ENSO) are present we must be careful
when assessing the return values since we must be convinced that the archive
covers a sufficiently long period to capture all the stages of the phenomenon.
As noted in \Sec{intro}, under such circumstances non-stationary techniques
employed on traditional time series and climate projections will be more
relevant. If on the other hand return values valid for the present period
are sought then ensemble forecasts are superior.

\section{Conclusion}
\label{sec:conclusion}
Return values estimated from long lead-time ensemble forecasts have been
investigated and found to yield good results. The immediate advantage is clear;
a huge data set of forecasts are readily available from the ECMWF archive.
The method yields return values of significant wave height that are comparable
to what is found from NORA10 but significantly higher than what is found from
ERA-I and ERA-40. This result was not totally unexpected since it is known
that ERA-I and especially ERA-40 tend to underestimate the upper percentiles
of the wave height distribution. The EPS estimates are probably too low
in enclosed seas (see \Fig{comet}).  Although we have only investigated
the extremes in the North Atlantic, the North Sea and the Norwegian Sea,
it appears likely that the extreme value estimates found from ERA-40 and
ERA-I are too low globally (as discussed by \citealt{caires05b,sterl05}
in the case of ERA-40).  However, the return value estimates from NORA10
\citep{aar12} and the present findings suggest that the corrected ERA-40
return estimates reported by \cite{caires05b} and \cite{sterl05} are too high.

Return value estimation from large ensembles at advanced lead times is a
general method which should be applicable to a wide range of atmospheric
and oceanographic variables if the conditions discussed in \Sec{method}
and \Sec{discussion} are met.  It is clear that the EPS archive represents
an unused resource which complements and perhaps yields more precise return
values than traditional reanalyses and hindcasts.

\section*{Acknowledgment} 
This work has been supported by the Research Council of Norway through
the project ``Wave Ensemble Prediction for Offshore Operations'' (WEPO,
grant no 200641) and through the European Union FP7 project MyWave (grant
no 284455). This study has also been part of a PhD program partially funded
by the Norwegian Centre for Offshore Wind Energy (NORCOWE) for OJA. The
Norwegian Deepwater Programme (NDP) financed the construction of the NORA10
hindcast archive.  The patient advice of Saleh Abdalla, Peter Janssen, Hans
Hersbach and Martin Leutbecher is greatly appreciated. We would also like
to thank the three anonymous reviewers. Their constructive comments helped
improve the manuscript.

{\clearpage}
\bibliography{/home/rd/diob/Doc/TeX/Bibtex/BreivikAbb,/home/rd/diob/Doc/TeX/Bibtex/Breivik} \newpage

%
\end{document}